\documentclass[english,aps:twocolumns]{revtex4}
\usepackage[T1]{fontenc}
\usepackage[latin9]{inputenc}
\usepackage{xcolor}
\usepackage{amsmath}
\usepackage{amsthm}
\usepackage{amssymb}
\usepackage{graphicx}
\PassOptionsToPackage{normalem}{ulem}
\usepackage{ulem}

\makeatletter

\providecolor{lyxadded}{rgb}{0,0,1}
\providecolor{lyxdeleted}{rgb}{1,0,0}

\@ifundefined{textcolor}{}
{%
 \definecolor{BLACK}{gray}{0}
 \definecolor{WHITE}{gray}{1}
 \definecolor{RED}{rgb}{1,0,0}
 \definecolor{GREEN}{rgb}{0,1,0}
 \definecolor{BLUE}{rgb}{0,0,1}
 \definecolor{CYAN}{cmyk}{1,0,0,0}
 \definecolor{MAGENTA}{cmyk}{0,1,0,0}
 \definecolor{YELLOW}{cmyk}{0,0,1,0}
}

\newcommand{\be }{\begin {equation}} \newcommand{\ee }{\end {equation}}

\newcommand{\ket}[1]{|#1\rangle}
\newcommand{\bra}[1]{\langle #1|}

\usepackage{subfigure}

\usepackage{enumitem}

\usepackage{babel}

\makeatother

\usepackage{babel}
\begin{document}

\title{Sequences of lower bounds for entropic uncertainty relations from
bistochastic maps}

\author{Paolo Giorda}
\email{magpaolo16@gmail.com}

\affiliation{Dip. Fisica , University of Pavia, via Bassi 6, I-27100 Pavia, Italy;
Consorzio Nazionale Interuniversitario per la Scienze fisiche della
Materia (CNISM), Italy}
\begin{abstract}
Given two orthornormal bases $\mathcal{A}$ and $\mathcal{B}$, the
basic form of the entropic uncertainty principle is stated in terms
of the sum of the Shannon entropies of the probabilities of measuring
$\mathcal{A}$ and $\mathcal{B}$ onto a given quantum state. State
independent lower bounds for this sum encapsulate the degree of incompatibility
of the observables diagonal in the $\mathcal{A}$ and $\mathcal{B}$
bases, and are usually derived by extracting as much information as
possible from the unitary operator $U$ connecting the two bases.
Here we show a strategy to derive sequences of lower bounds based
on alternating sequences of measurements onto $\mathcal{A}$ and $\mathcal{B}$.
The problem can be mapped into the multiple application of bistochastic
processes that can be described by the powers of the unistochastic
matrices directly derivable from $U$. By means of several examples
we study the applicability of the method. The results obtained show
that the strategy can allow for an advantage both in the pure state
and in the mixed state scenario. The sequence of lower bounds is obtained
with resources which are polynomial in the dimension of the underlying
Hilbert space, and it is thus suitable for studying high dimensional
cases.
\end{abstract}
\maketitle

\section{Introduction}

Uncertainty relations are some of the ways in which the peculiar behavior
of quantum systems with respect to classical ones is characterized.
The same feature described by these relation, the uncertainty associated
with the measurements results of distinct incompatible observables,
can be casted in several different ways depending on the context and
the aim. Originally, the product of two observables' variances were
used \cite{Robertson_Uncertainty_Principle,URHistory}, more recently
the relations have been stated in terms of sum of variances \cite{Sum_Variances_Collective}
or, when the spectrum of the observables is not relevant, in terms
of sum of entropic quantities \cite{Hystory_EURs_Collective,Deutsch_EURs}.
For discrete observables acting on finite dimensional Hilbert spaces
$\mathcal{H}_{M},\ M=dim\mathcal{H}_{M}$, the elementary form of
entropic uncertainty relations (EURs) is stated for two bases $\mathcal{A}=\left\{ \ket{a_{n}}\right\} _{n=1}^{M}$
and $\mathcal{B}=\left\{ \ket{b_{n}}\right\} _{n=1}^{M}$ as
\begin{eqnarray}
H\left(\mathcal{A},\mathcal{B}\right) & = & H\left(\mathcal{A}\right)+H\left(\mathcal{B}\right)\ge\mathcal{L}_{B}\label{eq: EURs definition}
\end{eqnarray}
where $H\left(\mathcal{A}\right)=-\sum_{i}p_{i}^{a}\log p_{i}^{a}$
is the Shannon entropy of the probability vector $\bar{p}^{a}=\left(p_{1}^{a},..,p_{M}^{a}\right)$
with $p_{i}^{a}=Tr\left[\rho\ket{a_{i}}\bra{a_{i}}\right]$; and analogously
$H\left(\mathcal{B}\right)=-\sum_{j}p_{j}^{b}\log p_{j}^{b}$, with
$p_{j}^{b}=Tr\left[\rho\ket{b_{j}}\bra{b_{j}}\right]$. Here $\mathcal{L}_{B}\ge0$
is a positive constant that lower bounds the sum for a given set of
quantum states $\rho$. When $\mathcal{L}_{B}$ is a function of the
measurements $\mathcal{A},\mathcal{B}$ only, it is termed as state
independent and the relation (\ref{eq: EURs definition}) and it is
satisfied by all states $\rho\in\mathcal{B}\left(\mathcal{H}\right)$.
In such case $\mathcal{L}_{B}$ encodes the degree of incompatibility
of the observables $\mathcal{A},\mathcal{B}$, since it puts a limit
to the irreducible amount of uncertainty, as measured by the sum of
the Shannon entropies, of the experiments represented by $\mathcal{A}\mbox{ and }\mathcal{B}$.
If instead $\mathcal{L}_{B}$ also depends on the von Neumann entropy
of the set of states $S_{c}=\left\{ \rho|S(\rho)=c\right\} $, it
provides a state dependent lower bound for all states in $S_{c}$,
and it therefore encodes the degree of incompatibility of $\mathcal{A}$
and $\mathcal{B}$ in presence of a given fixed value of entropy.
Given the definition of coherence for a given basis $\mathcal{A}$
with respect to a state $\rho$ in terms of the relative entropy of
coherence $C_{\mathcal{A}}\left(\rho\right)=-S\left(\rho\right)+H\left(\mathcal{A}\right)$
\cite{CoherenceGeneral}, the EURs can also be formulated as 
\begin{eqnarray}
C_{\mathcal{A}}\left(\rho\right)+C_{\mathcal{B}}\left(\rho\right) & \ge & \mathcal{L}_{B}-2S\left(\rho\right)\label{eq: EURs vs sum of coherences}
\end{eqnarray}
If $\mathcal{L}_{B}-2S\left(\rho\right)>0$, the latter formula constitutes
a non trivial lower bound for the sum of coherences. This connection
between the EURs and the sum of coherences naturally provides an interpretation
of complementarity property expressed by the EURs: the complementarity
between $\mathcal{A}$ and $\mathcal{B}$, and thus the minimum uncertainty
for the two experiments represented by $\mathcal{A}$ and $\mathcal{B}$,
is rooted in minimum of the sum of their coherences. The formulation
(\ref{eq: EURs vs sum of coherences}) will be used in the following,
it reduces to (\ref{eq: EURs definition}) for pure states, and it
may find application for special cases of entropic uncertainties with
memory \cite{Berta2010_EURs_Mixed_States} (see below).

Since the first formulation of the uncertainty principle in terms
of entropic quantities \cite{Hystory_EURs_Collective,Deutsch_EURs,MU},
several methods have been developed to provide tighter bounds $\mathcal{L}_{B}$
for (\ref{eq: EURs definition}), for its generalizations to mixed
states \cite{Berta2010_EURs_Mixed_States}, for more than two bases
\cite{EURs_Multiple_Measurement_Collective} and for generalized measurements
\cite{EURs_POVMs} (an extensive collection of methods, results, applications
and citations can be found in the excellent recent review \cite{Wehner_EURs_RMP}).
In many of these approaches, the goal has been to extract as much
information as possible from the unitary operator $U$ connecting
the two bases i.e., $\ket{b_{n}}=U\ket{a_{n}},\ \forall n$. In the
original fundamental works \cite{Deutsch_EURs,MU}, the authors initially
gave a state independent lower bound in terms of the largest overlap
between the elements of $\mathcal{A}\mbox{ and }\mathcal{B}$; in
particular Maassen and Uffink provided the following lower bound $\mathcal{L}_{MU}=-\log s_{MU}$,
where $s_{MU}=\max_{i,j}\left|\bra{a_{i}}\left.b_{j}\right\rangle \right|^{2}$
is the largest modulus square element of $U$ in the $\mathcal{A}$
basis. Subsequent approaches have successfully managed to exploit
more of the information contained in $U$ i.e., they have provided
lower bounds that depends on two matrix elements $\left|\bra{a_{i}}U\ket{a_{j}}\right|$
\cite{ColesPianiLBsHighDim,Zyc_2014_Majorization} or sub-matrices
of $U$ \cite{Friedland_2013_Original_Majorization,Zyc_2013_Majorization,Zyc_2014_Majorization,Zyc_2018_Majorization,Patrovi_2011_Original_Majorization}. 

\begin{figure}[h]
\subfigure{\includegraphics[bb=0bp 0bp 794bp 360bp,scale=0.24]{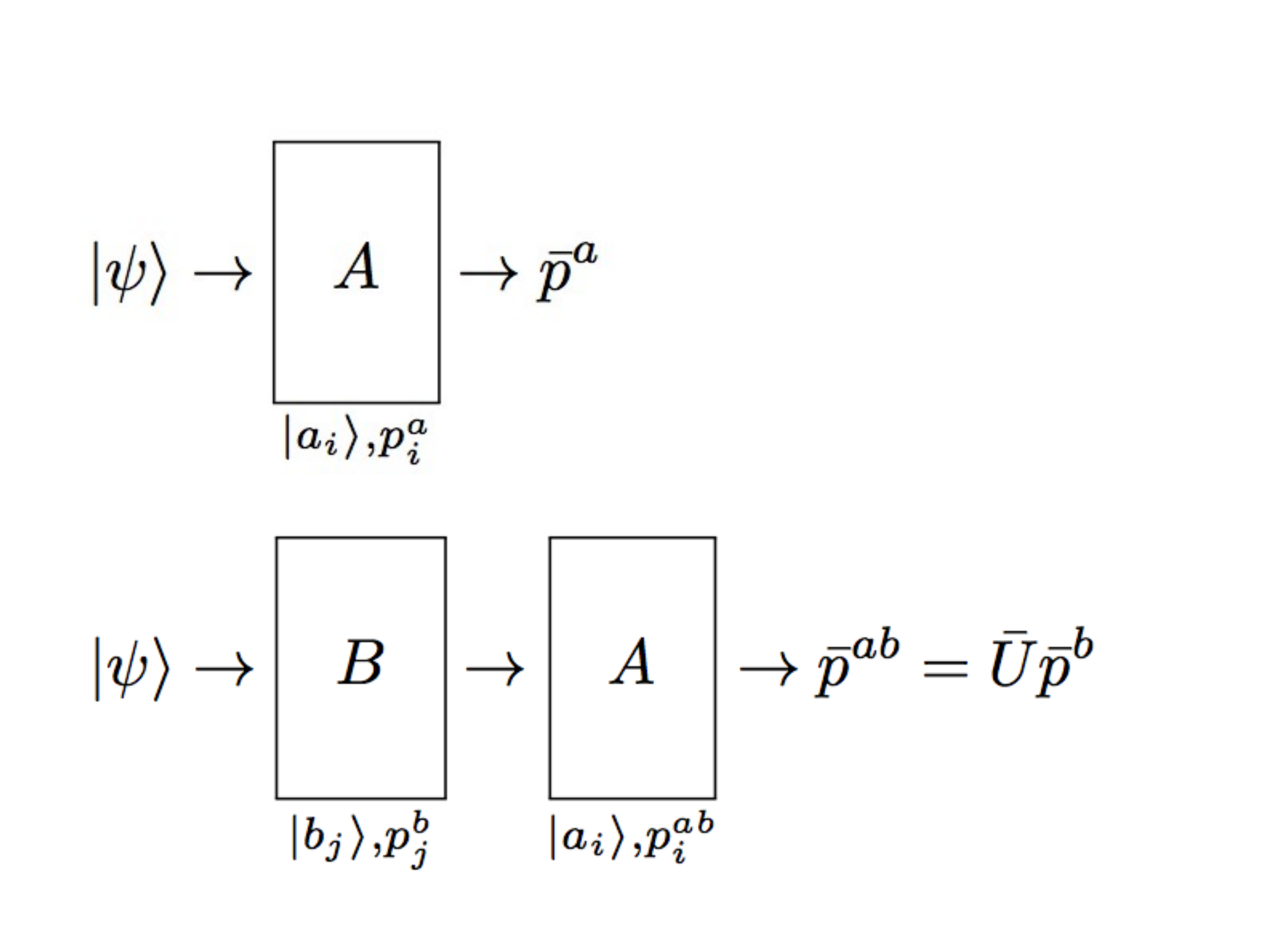}
} \subfigure{\includegraphics[bb=0bp 160bp 794bp 480bp,clip,scale=0.36]{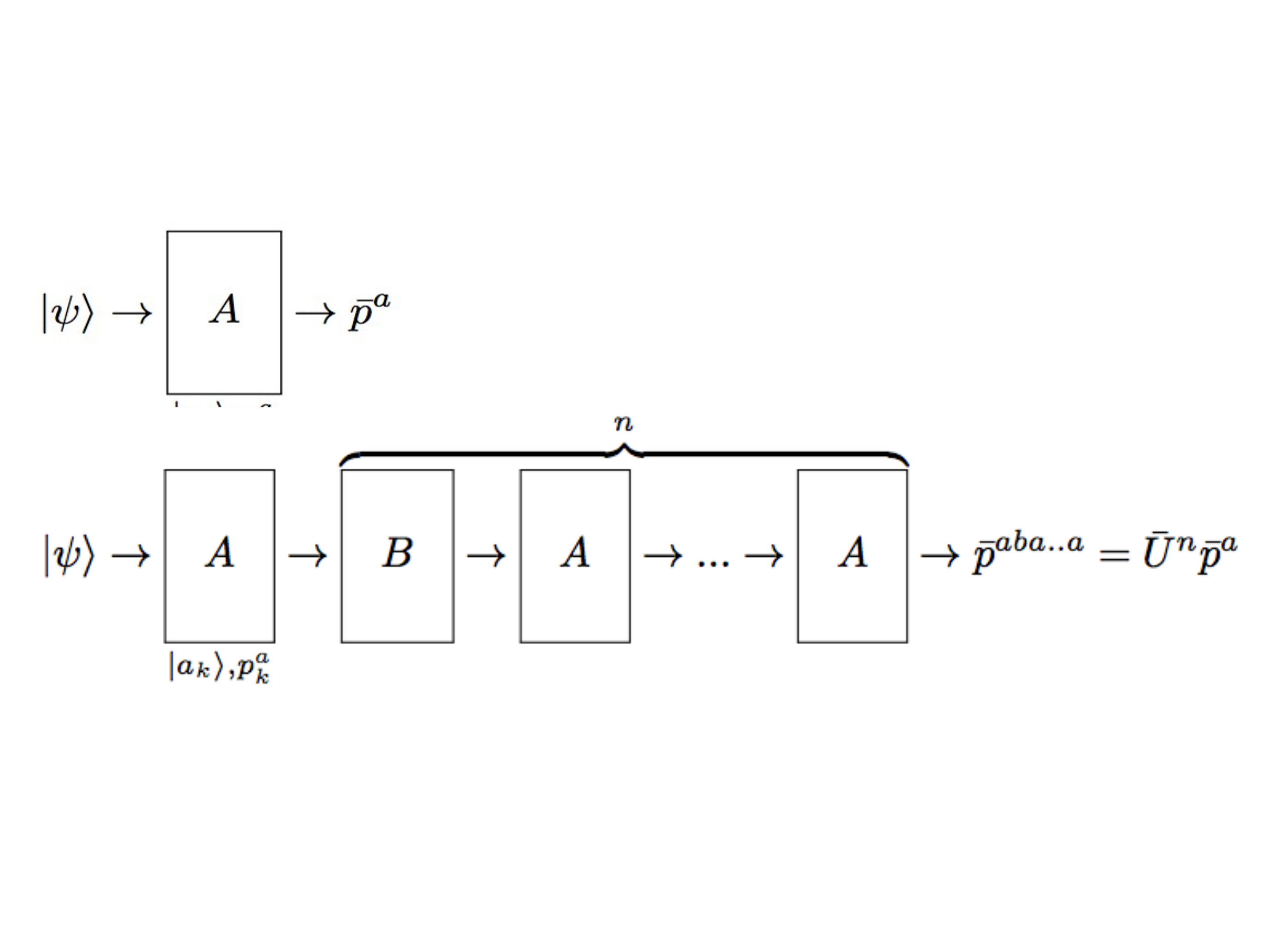}
}\caption{Measurement schemes at the basis of the method\textbf{ Left panel:
}example of one vs $n+1=2$ measurement scheme \textbf{Right panel:}
example of one vs $n+1$ measurements scheme when $\bar{U}=\bar{U}^{T}$.
In both cases, the final output probability vector pertaining the
last measurement stage $\mathcal{A}$ can be expressed in terms of
the $n$-th power of the unistochastinc matrix $\bar{U}$\label{fig: Measurement-schemes}}
\end{figure}

In the following we show how a basic strategy developed in \cite{Berta2010_EURs_Mixed_States,Coles_2012_L1_Bound}
can be extended (see Figure (\ref{fig: Measurement-schemes}) and
below) to alternating chains of $n+1$ measurements on $\mathcal{A}$
and $\mathcal{B}$ in order to provide a sequence of lower bounds
$\mathcal{L}_{n}$. We show that the problem of finding a tighter
lower bound can be mapped into finding the maximum element of a $n$-fold
product of bistochastic matrices whose factors are given by $\bar{U}$
and its transposed $\bar{U}^{T}$. Here $\bar{U}$ is the unistochastic
opeartor whose matrix elements $\bar{U}_{i,j}=\left|\bra{a_{i}}U\ket{a_{j}}\right|^{2}$
are given by the moduli squared elements of $U$. Indeed, the matrix
$\bar{U}$ and its transposed $\bar{U}^{T}$, being bi-stochastic,
can be seen as the realization of classical Markovian processes that
transform the input probability vectors $\bar{p}^{a}=\left(p_{1}^{a},..,p_{M}^{a}\right)^{T}$
and $\bar{p}^{b}=\left(p_{1}^{b},..,p_{M}^{b}\right)^{T}$ into output
ones i.e., $\bar{U}\bar{p}^{a}$ and $\bar{U}^{T}\bar{p}^{b}$ . As
we shall see, the multiple applications of such processes describe
the alternating sequences  of measurements on $\mathcal{A}$ and $\mathcal{B}$.
In the most simple case where $\bar{U}=\bar{U}^{T}$ is symmetric,
this amounts to implement the Markovian processes $\left(\bar{U}\right)^{n}\bar{p}^{a}$
and $\left(\bar{U}\right)^{n}\bar{p}^{b}$ . The effects of these
classical Markovian ``filters'' clearly depends on the existing
relation between $\mathcal{A}$ and $\mathcal{B}$, as described by
$\bar{U}$. By upper bounding the effects of such ``filters'' one
can then obtained the desired state independent and state dependent
lower bounds. Indeed, for any $n\ge1$, by essentially comparing the
output probabilities $\left(\bar{U}\right)^{n}\bar{p}^{a}$ and $\left(\bar{U}\right)^{n}\bar{p}^{b}$
with $\bar{p}^{a}$ or $\bar{p}^{b}$ by means of the classical relative
entropy, one can obtain a lower bound of the form 
\begin{eqnarray}
\mathcal{L}_{n} & = & \mathcal{U}_{n}+\mathcal{S}_{n}\label{eq: Ln definition}
\end{eqnarray}
where $\mathcal{U}_{n}$ is a state independent term and $\mathcal{S}_{n}$
is a term that solely depends on the von Neumann entropy of the state
$S(\rho)$. As we shall see the sequence $\mathcal{U}_{n},\ n>1$
provides new expressions for state independent lower bounds, that
in the simplest case $\bar{U}=\bar{U}^{T}$ can be written as 
\begin{eqnarray}
\mathcal{U}_{n} & = & \frac{-\log s_{n}}{n+1}\label{eq: Un definition}
\end{eqnarray}
where $s_{n}=\left[\max_{i,j}\left(\bar{U}^{n}\right)_{i,j}\right]^{2}$
is the squared largest matrix element in the $\mathcal{A}$ basis
of the bi-stochastic matrix $\bar{U}^{n}$. We shown that the $s_{n}$
can be easily computed for each $n$ and for arbitrary dimensions
in terms of the eigenvalues of $\bar{U}$, its eigenvectors and their
overlap with the some of the elements of $\mathcal{A}$. On the other
hand, the term dependent on the von Neumann entropy can be simply
written for each $n$ as 
\begin{eqnarray}
S_{n} & = & \frac{2n}{n+1}S\left(\rho\right)\label{eq: Sn definition}
\end{eqnarray}
For fixed bases $\mathcal{A}$,$\mathcal{B}$ and $U$, the tightest
expression of the lower bound provided by this method is given by
$\mathcal{L}=\max_{n}\mathcal{L}_{n}$. As we shall see the method
can also be easily applied when $\bar{U}\neq\bar{U}^{T}$. For $n=1$
one recovers the basic result \cite{Berta2010_EURs_Mixed_States}
\begin{eqnarray}
\mathcal{L}_{1} & = & -\log s_{MU}+S\left(\rho\right)\label{eq: L1 definition}
\end{eqnarray}
which extended the Maassen-Uffink result \cite{MU} to mixed states.
For $n>1$ the various state independent bounds $\mathcal{U}_{n}$
are shown to provide in some situations a significant improvement
with respect to (\ref{eq: L1 definition}) and to bounds obtained
with other strategies \cite{ColesPianiLBsHighDim,Zyc_2014_Majorization,deVincenteLBsHighDim}.
This is in particular true when the dimension $M$ of the underlying
Hilbert space is large. The improvement is maintained and in certain
cases enhanced by the $\mathcal{L}_{n}$'s when one considers mixed
states e.g., when one seeks for lower bound of the sum of coherences
(\ref{eq: EURs vs sum of coherences}) or conditional entropic quantities.
In this case, we show and example in which the lower bounds provided
by the $\mathcal{L}_{n}$ are the only non-trivia i.e., non-zero,
and simply computable ones.

In the following we first show how the main result (\ref{eq: Ln definition})
can be obtained (section \ref{sec: Two-observables-lower-bound-theory})
and discuss its interpretation. We then show how the terms $\mathcal{U}_{n}$
can be easily expressed in terms of the eigenvalues and eigenvectors
of $\bar{U},\bar{U}^{n}$ ,$\bar{U}\bar{U}^{T}$ and $\bar{U}^{T}\bar{U}$
(subsection \ref{subsec: Evaluation-of-the-lower-bounds}). In order
to test the above outlined strategy, we then apply it to different
examples and compare the results with those obtained with other strategies
(section \ref{sec: Examples}). We finally recap and discuss the results
obtained and give some perspective on the possible extensions of the
method proposed (section \ref{sec: Discussion}).

\section{Two observables lower bounds\label{sec: Two-observables-lower-bound-theory}}

In this section we describe how the sequence of lower bounds $\mathcal{L}_{n}$
can be derived. The various $\mathcal{L}_{n}$ are ultimately based
on the use of (quantum) relative entropies in order to compare the
output probability vectors of different experiments. We first show
the case $n=1$ (already obtained in \cite{Berta2010_EURs_Mixed_States,Coles_2012_L1_Bound,Wehner_EURs_RMP})
and later we extend it to the case $n>1$. For $n=1$, in the first
experiment one measures the observable defined by the basis $\mathcal{A}$
and obtains the output probability vector $\bar{p}^{a}=\left(p_{1}^{a},p_{2}^{a},..,p_{N}^{a}\right)$.
The second probability vector $\bar{p}^{ab}=\left(p_{1}^{ab},p_{2}^{ab},..,p_{N}^{ab}\right)$
is the output of a second experiment where first one measures $\rho$
onto the basis $\mathcal{B}$ and then applies a measurement in the
$\mathcal{A}$ basis. Here 
\begin{eqnarray*}
p_{i}^{ab} & = & \sum_{j}p_{j}^{b}\left|\bra{b_{j}}\left.a_{i}\right\rangle \right|^{2}
\end{eqnarray*}
is the probability of obtaining $\ket{a_{i}}$ after the $\mathcal{B}$
measurement process has occurred on $\rho$. The classical relative
entropy (Kullback\textendash Leibler divergence) 
\begin{eqnarray*}
D\left(\bar{p}^{a}||\bar{p}^{ab}\right) & = & -\sum_{i}p_{i}^{a}\log\frac{p_{i}^{a}}{p_{i}^{ab}}
\end{eqnarray*}
is a measure of the statistical distinguishability between the two
probability vectors $\bar{p}^{a},\bar{p}^{ab}$. In this case it measures
how much the measurement process $\mathcal{A}$ is ``disturbed''
when one first measures $\rho$ onto the observable defined by $\mathcal{B}$.
The disturbance clearly depends on the relation between the two basis
$\mathcal{A},\mathbf{\mathcal{B}}$. Analogously, by exchanging the
role of $\mathcal{A}$ and $\mathbf{\mathcal{B}}$, one can use $D\left(\bar{p}^{b}||\bar{p}^{ba}\right)$
to compare the output of two different measurement process where in
the first one directly measures $\mathcal{B}$ on $\rho$, while in
the second one first measures $\mathcal{A}$ and then $\mathcal{B}$.
In order to obtain the lower bound, one then introduces the maps $A\left(\cdot\right)=\sum\Pi_{i}^{a}\left(\cdot\right)\Pi_{i}^{a}$
and $B\left(\cdot\right)=\sum\Pi_{j}^{b}\left(\cdot\right)\Pi_{j}^{b}$,
that describe the measurement processes on $\mathcal{A}$ and $\mathbf{\mathcal{B}}$
respectively; here $\Pi_{i}^{a}=\ket{a_{i}}\bra{a_{i}}$ and $\Pi_{j}^{b}=\ket{b_{j}}\bra{b_{j}}$.
One can thus write the relative entropy $D\left(\bar{p}^{a}||\bar{p}^{ab}\right)=S\left(A\left(\rho\right)||A\cdot B\left(\rho\right)\right)\equiv S\left(A||AB\right)$
in terms of the quantum relative entropy $S\left(A||AB\right)$ \cite{Wehner_EURs_RMP}
between the states $A\left(\rho\right)=\sum\Pi_{i}^{a}p_{i}^{a}$
and $A\circ B\left(\rho\right)=\sum\Pi_{i}^{a}p_{i}^{ab}$; analogously
$D\left(\bar{p}^{b}||\bar{p}^{ba}\right)\equiv S\left(B||BA\right)$.
Since $A,B$ are completely positive maps one can then use the data
process inequality \cite{Lieb DataProcessInequality,NielsenChuangBook,Wehner_EURs_RMP}
to write 
\begin{eqnarray*}
-S\left(\rho\right)+S(B\left(\rho\right))=S\left(\rho||B\right) & \ge & S\left(A||AB\right)
\end{eqnarray*}
where now $S(B\left(\rho\right))=H\left(\mathcal{B}\right)$ i.e.,
the von Neumann entropy of the state $B\left(\rho\right)$ coincides
with the Shannon entropy of the probability vector $\bar{p}^{b}$.
By exchanging the role of $\mathcal{A}$ and $\mathbf{\mathcal{B}}$
one obtains the analogous relation $-S\left(\rho\right)+H\left(\mathcal{A}\right)=S\left(\rho||A\right)\ge S\left(B||BA\right)$.
Summing up the two relations one has 
\begin{eqnarray}
H\left(\mathcal{A}\right)+H\left(\mathcal{B}\right) & \ge & S\left(A||AB\right)+S\left(B||BA\right)+2S\left(\rho\right)\label{eq:LB in terms of QRelative entropy n=00003D1}
\end{eqnarray}
This is the prototypical expression that will allow us to obtain the
lower bounds $\mathcal{L}_{n}$ for any $n>1$. Since now $S\left(A||AB\right)=D\left(\bar{p}^{a}||\bar{p}^{ab}\right)=-H\left(\mathcal{A}\right)+C\left(A||AB\right)$,
where $C\left(A||AB\right)=-\sum_{i}p_{i}^{a}\log p_{i}^{ab}$ is
the cross-entropy \cite{NielsenChuangBook} between $\bar{p}^{a}$
and $\bar{p}^{ba}$, one finally obtains 
\begin{eqnarray}
H\left(\mathcal{A}\right)+H\left(\mathcal{B}\right) & \ge & \frac{C\left(A||AB\right)+C\left(B||BA\right)}{2}+S\left(\rho\right)\label{eq: LB in terms of Cross entropy n=00003D1}
\end{eqnarray}
Since $\forall i,j$ the probabilities $p_{i}^{ab},p_{j}^{ba}\le\max_{h,k}\bar{U}_{hk}=s_{MU}$,
then $C\left(A||BA\right),C\left(B||AB\right)\ge-\log s_{MU}$, and
one obtains $\mathcal{L}_{1}$ (\ref{eq: L1 definition}) in terms
of the Maassen-Uffink state independent lower bound. The above strategy
can be extended to more complex measurement processes. Indeed the
bound $\mathcal{L}_{2}$ can be obtained by comparing the following
different processes, see also Figure \ref{fig: Measurement-schemes}-left
panel. The first, described by $\bar{p}^{a}$ is the measurement $A\left(\rho\right)$
on $\mathcal{A}$; the second is the sequence of measurements $A\circ B\circ A\left(\rho\right)$
described by the probability vector $\bar{p}^{aba}$:
\begin{eqnarray}
p_{i}^{aba} & = & \sum_{k}p_{k}^{a}\sum_{t}\left|\bra{a_{k}}\left.b_{t}\right\rangle \right|^{2}\left|\bra{b_{t}}\left.a_{i}\right\rangle \right|^{2}\label{eq: probability ABA}
\end{eqnarray}
the (quantum) relative entropy $D\left(\bar{p}^{a}||\bar{p}^{aba}\right)=S\left(A||ABA\right)$
now measures the statistical distinguishability between the probability
of measuring $\mathcal{A}$ directly on $\rho$ or onto $B\circ A\left(\rho\right)$.
It thus is a measure of how much the interposition of a measurement
stage $\mathcal{B}$ between the two $\mathcal{A}$ stages of measurements
changes the initial $\mathcal{A}$ outcome probability vector $\bar{p}^{a}$.
In other words it is a measure of the effect introduced by interposing
a $\mathcal{B}$ ``filtering'' operation between two $\mathcal{A}$
consecutive measurements processes. The same strategy can be applied
by exchanging the role of $\mathcal{A},\mathbf{\mathcal{B}}$, thus
obtaining the vector $\bar{p}^{bab}$ and $D\left(\bar{p}^{b}||\bar{p}^{bab}\right)=S\left(B||BAB\right)$.
In order to obtain a bound analogous to (\ref{eq:LB in terms of QRelative entropy n=00003D1}),
one makes a repeated use of the data process inequalities; indeed
\begin{eqnarray*}
S\left(A||ABA\right) & \le & S\left(\rho||BA\right)=-S\left(\rho\right)+H(\mathcal{B})+S\left(B||BA\right)\le\\
 & \le & -S\left(\rho\right)+H(\mathcal{B})+S\left(\rho||A\right)=\\
 & = & -2S\left(\rho\right)+H(\mathcal{A})+H(\mathcal{B})
\end{eqnarray*}
Applying the same arguments to $S\left(B||BAB\right)$ and summing
the two results one thus obtains 
\begin{eqnarray}
H\left(\mathcal{A}\right)+H\left(\mathcal{B}\right) & \ge & \frac{S\left(A||ABA\right)+S\left(B||BAB\right)}{2}+2S\left(\rho\right)\label{eq: LB in terms of QRelative entropy n=00003D2}
\end{eqnarray}
The same reasonings can then be applied to an arbitrary sequence of
alternating measurement processes $ABA...$ and $BAB...$., see also
Figure \ref{fig: Measurement-schemes}-right panel for the sequence
we explicitly consider below. We now sketch the proof for obtaining
the $n$-th order lower bound. At the at the $n$-th order one needs
to consider: a measurement process in which one measures $\mathcal{A}$
($\mathcal{B}$) directly on $\rho$; a sequences of alternating measurement
processes that terminates again with a measurement on $\mathcal{A}$
($\mathcal{B}$). Suppose now $n$ is even, then the two sequences
one has to consider will be
\begin{eqnarray*}
A\overbrace{BA...BA} & \mbox{and } & B\overbrace{AB...AB}
\end{eqnarray*}
where the number of operators underneath each bracket is $n$. Suppose
now we start with the first sequence, the first two step of the process
will give the following chain of inequalities 
\begin{eqnarray*}
S\left(A||ABA...BA\right) & \le & S\left(\rho||BA...BA\right)=-S\left(\rho\right)+H(\mathcal{B})+S\left(B||BA...ABA\right)\le\\
 & \le & -S\left(\rho\right)+H(\mathcal{B})+S\left(\rho||ABA...A\right)=\\
 & = & -2S\left(\rho\right)+H(\mathcal{A})+H(\mathcal{B})+S\left(A||ABA...A\right)\le\\
 & \le & ....
\end{eqnarray*}
Thus at each step, by writing $S\left(\rho||C.....A\right)=-S\left(\rho\right)+H(\mathcal{C})+S\left(C||C.....A\right)$
, with $C$ being equal to $A$ or $B$ depending on step, one gets
two new addendi $-S\left(\rho\right)$ and $H(C)$. At the last of
the $n$ steps of the process for the given sequence one finds $S\left(\rho||A\right)=-S\left(\rho\right)+H(A)$.
Therefore, since $n$ is even, the expression at the end of the chain
of inequalities is $-nS\left(\rho\right)+\frac{n}{2}\left(H(\mathcal{A})+H(\mathcal{B})\right)$
. The same result is obtained if one starts with the other sequence
$BAB...AB$. Thus, by summing up the two results and dividing by $n$
one gets
\begin{eqnarray}
H\left(\mathcal{A}\right)+H\left(\mathcal{B}\right) & \ge & \frac{S\left(A||ABA..A\right)+S\left(B||BAB...B\right)}{n}+2S\left(\rho\right)\label{eq: LB in terms of QRelative entropy n generic}
\end{eqnarray}
Analogously, it's easy to check that when $n$ is odd the final expressions
obtained for the two sequences are $-nS\left(\rho\right)+\frac{n+1}{2}H(\mathcal{A})+\frac{n-1}{2}H(\mathcal{B})$
and $-nS\left(\rho\right)+\frac{n-1}{2}H(\mathcal{A})+\frac{n+1}{2}H(\mathcal{B})$
respectively. Thus, by summing them up and dividing by $n$ one again
obtains (\ref{eq: LB in terms of QRelative entropy n generic}). The
latter expression allows now to obtain a lower bound simply by expressing
the relative entropies $S\left(A||ABA..\right)=-H\left(\mathcal{A}\right)+C\left(A||ABA..\right)$
in terms of the cross entropies $C\left(A||ABA..\right)$. By multiplying
by $n$ both sides of the inequality (\ref{eq: LB in terms of QRelative entropy n generic})
and collecting the terms $H\left(\mathcal{A}\right)+H\left(\mathcal{B}\right)$
one then has 
\begin{eqnarray}
H\left(\mathcal{A}\right)+H\left(\mathcal{B}\right) & \ge & \frac{C\left(A||ABA..A\right)+C\left(B||BAB...B\right)}{n+1}+\frac{2n}{n+1}S\left(\rho\right)\label{eq: LB in terms of Cross entropy n generic}
\end{eqnarray}
The above formula allows to obtain the lower bounds $\mathcal{L}_{n}$
for any $n>1$ in a compact way. We show how by starting from $\mathcal{L}_{2}$.
The probabilities $\bar{p}_{i}^{aba}$ (\ref{eq: probability ABA})
and $\bar{p}_{j}^{bab}$ can be written in terms of the matrix elements
of the product $\bar{U}\bar{U}^{T}$ and $\bar{U}^{T}\bar{U}$ respectively.
Indeed, $\bar{p}_{i}^{aba}$ (\ref{eq: probability BAB}) is written
in terms of the elements 
\begin{eqnarray*}
\sum_{t}\left|\bra{a_{k}}\left.b_{t}\right\rangle \right|^{2}\left|\bra{b_{t}}\left.a_{i}\right\rangle \right|^{2} & = & \sum_{t}\left|\bra{a_{k}}U\ket{a_{t}}\right|^{2}\left|\bra{a_{t}}U^{\dagger}\ket{a_{i}}\right|^{2}\\
 &  & \sum_{t}\left|U_{kt}\right|^{2}\left|U_{ti}^{\dagger}\right|^{2}=\\
 & = & \bra{a_{k}}\bar{U}\bar{U}^{T}\ket{a_{i}}
\end{eqnarray*}
and analogously 
\begin{eqnarray}
p_{j}^{bab} & = & \sum_{k}p_{k}^{b}\sum_{t}\left|\bra{b_{k}}\left.a_{t}\right\rangle \right|^{2}\left|\bra{b_{j}}\left.a_{i}\right\rangle \right|^{2}=\nonumber \\
 & = & \sum_{k}p_{k}^{b}\bra{a_{k}}\bar{U}^{T}\bar{U}\ket{a_{j}}\label{eq: probability BAB}
\end{eqnarray}
We now observe that since $\bar{U}$ is bi-stochastic, $\bar{U}^{T}$
and $\bar{U}^{T}\bar{U},\bar{\ U}\bar{U}^{T}$ are bi-stochastic too
\cite{ZycUnistochasticMtx}. The lower bound $\mathcal{L}_{2}$ can
then be written in terms of the largest matrix elements 
\begin{eqnarray*}
s_{2} & = & \left(\max_{k,i}\bra{a_{k}}\bar{U}\bar{U}^{T}\ket{a_{i}}\right)\left(\max_{k,j}\bra{a_{k}}\bar{U}^{T}\bar{U}\ket{a_{j}}\right)
\end{eqnarray*}
Since $C\left(A||ABA\right)+C\left(B||BAB\right)\ge-\log s_{2}$,
using (\ref{eq: LB in terms of Cross entropy n generic}) for $n=2$
one has
\begin{eqnarray}
H\left(\mathcal{A}\right)+H\left(\mathcal{B}\right) & \ge & \mathcal{L}_{2}=-\frac{1}{3}\log s_{2}+\frac{4}{3}S\left(\rho\right)\label{eq: LB Bn n =00003D 2}
\end{eqnarray}
The same strategy can be straightforwardly applied to higher order
processes. One obtains the sequence of state independents lower bounds
$\{\mathcal{L}_{n}\}$ where for each $n$
\begin{eqnarray*}
H\left(\mathcal{A}\right)+H\left(\mathcal{B}\right) & \ge & \mathcal{L}_{n}=\mathcal{U}_{n}+\mathcal{S}_{n}=\\
 & = & -\frac{\log s_{n}}{n+1}+\frac{2n}{n+1}S\left(\rho\right)
\end{eqnarray*}
The state independent part of the bound $\mathcal{U}_{n}$ depends
on the relation between $\mathcal{A},\mathbf{\mathcal{B}}$ through
\begin{eqnarray*}
s_{n} & = & \left(\max_{k,i}\bra{a_{k}}\bar{U}\bar{U}^{T}\bar{U}\bar{U}^{T}..\ket{a_{i}}\right)\left(\max_{k,i}\bra{a_{k}}\bar{U}^{T}\bar{U}\bar{U}^{T}\bar{U}..\ket{a_{i}}\right)
\end{eqnarray*}
i.e., the largest element of the n-fold matrix product $\bar{U}\bar{U}^{T}\bar{U}\bar{U}^{T}..$.
or $\bar{U}^{T}\bar{U}\bar{U}^{T}\bar{U}..$.. If now $\bar{U}=\bar{U}^{T}$
is symmetric then the expression simplifies to $s_{n}=\left(\max_{i,j}\bra{a_{k}}\bar{U}^{n}\ket{a_{i}}\right)^{2}$.
The term $\mathcal{S}_{n}$ instead only depends on the von Neumann
entropy of the state $\rho$. Overall, given two bases $\mathcal{A},\mathbf{\mathcal{B}}$
and for fixed level of entropy $S\left(\rho\right)$, the best lower
bound provided by the above described strategy is given by $\mathcal{L}=\max_{n}\mathcal{L}_{n}$. 

\subsection{Discussion of the results: general considerations.}

Before analyzing the method that one can use to evaluate $s_{n}$
in a simple way, we first comment on some properties of the found
bounds that can be understood without explicitly computing them. We
start with the $\mathcal{S}_{n}$ part. If $\mathcal{A}\equiv\mathcal{B}$,
then $\bar{U}=\mathbb{I}$, $s_{n}=1,\ \forall n$ and the minimum
for the sum of entropies is given by twice the von Neumann entropy
of the state. Indeed, in this case $\mathcal{L}_{n}=\mathcal{S}_{n}\ \forall n$
and, since $\mathcal{S}_{n+1}\ge\mathcal{S}_{n}$, the best lower
bound in the sequence is provided by $\mathcal{L}_{n\rightarrow\infty}=\mathcal{S}_{n\rightarrow\infty}=2S\left(\rho\right)$.
This term captures the obvious feature that for states with a given
fixed von Neuman entropy and whatever the relation between $\mathcal{A}\mbox{ and }\mathcal{B}$,
it holds $H\left(\mathcal{A}\right),H\left(\mathcal{B}\right)\ge S\left(\rho\right)$
and thus $H\left(\mathcal{A}\right)+H\left(\mathcal{B}\right)\ge2S\left(\rho\right)$.
Since $\forall n,\ \mathcal{S}_{n}>S\left(\rho\right)$, we may expect
that the bounds given by the $\mathcal{L}_{n}$'s may provide an improvement
with respect to other existing bounds, in particular $\mathcal{L}_{1}$,
when the entropic part plays a relevant role; which is for example
the case of conditional entropic uncertainty relations and sum of
coherences (see below).

As for the state independent $\mathcal{U}_{n}$ part, the latter provide
a state independent bound that is usable also for pure states. In
the next section we will see how the $\mathcal{U}_{n}$ depend on
the properties (eigenvalues) of $\bar{U}$. Here, in order to discuss
the $\mathcal{U}_{n}$'s physical meaning, we first discuss the case
$n=2$. The probabilities $p_{i}^{aba}$ can be written in terms of
initial probability vector $\bar{p}^{a}$ and observing that $p_{i}^{aba}$
is the $i$-th element of the vector probability $\bar{p}^{aba}=\bar{U}^{T}\bar{U}\bar{p}^{a}$.
Therefore $\bar{p}^{aba}$ is the output probability distribution
of the classical bi-stochastic process modeled by $\bar{U}\bar{U}^{T}$,
acting onto initial vector probability vectors $\bar{p}^{a}$; the
same argument also holds for $\bar{p}^{bab}$, $\bar{p}^{b}$ and
$\bar{U}\bar{U}^{T}$. Therefore $\bar{U}\bar{U}^{T}$ and $\bar{U}^{T}\bar{U}$
act as classical ``filters'' that modify the initial probability
distributions $\bar{p}^{a},\bar{p}^{b}$. Then $s_{2}$ represents
the upper-bound to the largest probability of obtaining the outputs
$\ket{a_{i}}$ and $\ket{b_{j}}$ by means of the bi-stochastic processes
modeled by $\bar{U}\bar{U}^{T}$ $\left(\bar{U}^{T}\bar{U}\right)$.
Analogous considerations can be applied at any order $n$, and the
effect of the $n$-th order filters depends on the relation between
the two bases $\mathcal{A},\mathbf{\mathcal{B}}$ as described by
the unistochastic operator $\bar{U}$. \\ Since $\bar{p}^{abab...}$
and $\bar{p}^{baba...}$ are the result of the application of a sequence
of $n$ bi-stochastic processes to the initial probability vectors
$\bar{p}^{a},\bar{p}^{b}$ , with increasing $n$ the effect is to
produce an increasing level of mixing. Being $s_{n}$ composed by
the product of upper bounds to $p_{i}^{abab...}$ and $p_{j}^{baba...}$
respectively, it's information content is thus related to the amount
of global mixing introduced by the $n$-steps bi-stochastic processes
modeled by $\bar{U}\bar{U}^{T}\bar{U}\bar{U}^{T}..$. and $\bar{U}^{T}\bar{U}\bar{U}^{T}\bar{U}..$..
. Thus one can expect that if $m>n$ then $s_{m}<s_{n}$ and $-\log s_{m}>-\log s_{n}$;
this however in general does not imply that $\mathcal{U}_{m}>\mathcal{U}_{n}$
since 
\begin{eqnarray}
\frac{\mathcal{U}_{m}}{\mathcal{U}_{n}} & = & \frac{-\log s_{m}}{-\log s_{n}}\left(\frac{n+1}{m+1}\right)\label{eq: Ration Um Un}
\end{eqnarray}
and $\frac{n+1}{m+1}<1$. Aside from simple analytic examples (see
the qubit case below), finding general conditions for determining
whether and when $\mathcal{U}_{m}>\mathcal{U}_{n}$ seems difficult
in the general case, and one must to resort to study case by case
or check numerically. In the examples we provide below we show that
indeed with growing $n$ one can improve the overall bound. The largest
instance of $n>1$ that one has to check is however limited, since
the above reasoning lead to expect that for large $n$ the level of
mixing introduced by the product of bistochastic maps e.g. $\bar{U}\bar{U}^{T}\bar{U}\bar{U}^{T}..$
reaches its maximum i.e., $p_{i}^{abab...},p_{j}^{baba...}\approxeq1/M$
and $\lim_{n\rightarrow\infty}\mathcal{U}_{n}\approx\lim_{n\rightarrow\infty}-2\log M/\left(n+1\right)=0$.
Therefore, independently on the existing relation between $\mathcal{A},\mathbf{\mathcal{B}}$,
for large $n$ the lower bound $\mathcal{U}_{n}\rightarrow0$. This
consideration clearly limits the number of lower bounds $\mathcal{U}_{n}$
that one needs to evaluate to determine the best bound $\mathcal{L}$.
This fact is in particular true for the situations in which $\mathcal{A},\mathbf{\mathcal{B}}$
are very close to being mutually unbiased, since in this case one
should expect that $\bar{U}^{T}\approx\bar{U}\approx U^{*}$ where
$U^{*}$ is the van der Waerden matrix \cite{ZycUnistochasticMtx}
i.e., the matrix whose elements are all equal to $1/M$. If this is
the case then one expects that for pure states the best lower bound
is provided by the Maassen-Uffink result $\mathcal{U}_{1}$ since
$\forall n\ s_{n}\approx1/M$, $\mathcal{U}_{1}>\mathcal{U}_{n}$,
and $\mathcal{U}_{1}$ as expected provides a tight lower bound. However,
if $\mathcal{A},\mathbf{\mathcal{B}}$ are sufficiently far from being
mutually unbiased the bound will be given by $\mathcal{U}_{n}$, $n>1$
, and in the case of mixed states case, by $\mathcal{L}_{n}$, $n>1$.
These general considerations will become clearer in the next sections,
where we analyze in detail the state independent part $\mathcal{U}_{n}$,
and we discuss some examples.

\subsection{Evaluation of the state independent lower bounds $U_{n}$\label{subsec: Evaluation-of-the-lower-bounds}}

To evaluate the lower bounds $\mathcal{U}_{n}$ one needs in principle
to evaluate the maximum matrix element of n-fold products of the kind
$\bar{U}\bar{U}^{T}\bar{U}\bar{U}^{T}..$. for $n>1$. This can become
a demanding computational task, especially when the dimension $M$
of the given Hilbert space is large. However, on the one hand, the
computation complexity is that of matrix multiplication and thus is
polynomial in the dimension $M$. On the other hand, by taking advantage
of the properties of the matrix $\bar{U}$ one can drastically reduce
the complexity, since the $s_{n}$ can be written in a compact form
that depends on the eigenvalues of $\bar{U}$. We start by analyzing
the simplest scenario i.e., when $\bar{U}^{T}=\bar{U}$ is symmetric
and has the properties detailed below. Since $\bar{U}$ is real and
symmetric it can be diagonalized by means of a real orthogonal matrix
$O$ such that $OD_{\bar{U}}O^{T}$ and $D_{\bar{U}}=diag\left(\bar{u}_{1},\bar{u}_{2},..,\bar{u}_{N}\right)$
where $\bar{u}_{i}$ are the eigenvalues of $\bar{U}$ in decreasing
order; since $\bar{U}$ is bi-stochastic one has that: the maximal
eigenvalue is $\bar{u}_{1}=1$ and it corresponds to the uniform normalized
eigenvector $\ket{u_{1}}=\sum_{i}\ket{a_{i}}/\sqrt{M}$ \cite{ZycUnistochasticMtx};
furthermore $\left|\bar{u}_{i}\right|\le1,\ \forall i>1$. In the
following we suppose for simplicity that that $1>\bar{u}_{i}\ge0,\ \forall i>1$
(some other special cases, in particular when $\bar{U}$ is not symmetric,
are treated in Appendix \ref{sec: Appendix Evaluation of L_n, Ubar non symmetric}).
Suppose now that at the first order $n=1$ the largest matrix element
of $\bar{U}$ is given by $\bra{a_{i}}\bar{U}\ket{a_{j}}$ for some
specific pair $\left(i,j\right)$, then $\sqrt{s_{1}}=\sum_{k}\bar{u}_{k}\bra{a_{i}}\left.\bar{u}_{k}\right\rangle \bra{\bar{u}_{k}}\left.a_{j}\right\rangle $.
Since $\bar{U}^{n}=\sum_{k}\bar{u}_{k}^{n}\ket{\bar{u}_{k}}\bra{\bar{u}_{k}}$
is diagonal in the same basis for all $n$, and since for all $k$
and for all $m\ge n$ one has that $\bar{u}_{k}^{n}\ge\bar{u}_{k}^{m}$,
it obviously follows that at any order $n$ the lower bound will be
given by the same matrix element $i,j$ i.e., $\sqrt{s_{n}}=\bra{a_{i}}\bar{U^{n}}\ket{a_{j}}=\sum_{k}\bar{u}_{k}^{n}\bra{a_{i}}\left.\bar{u}_{k}\right\rangle \bra{\bar{u}_{k}}\left.a_{j}\right\rangle $.
Therefore, in order to evaluate $s_{n}$ one only needs to: $i)$
determine the pair $\left(i,j\right)$ that identifies $\bar{U}$'s
largest matrix element $ii)$ find the eigenvalues and eigenvectors
of $\bar{U}$ $iii)$ evaluate the coefficients $U_{i,j}^{k}=\bra{a_{i}}\left.\bar{u}_{k}\right\rangle \bra{\bar{u}_{k}}\left.a_{j}\right\rangle ,\ \forall k$.
Then the lower bounds $\mathcal{U}_{n}$ can then be easily computed
in a compact form since $\forall n$ 
\begin{eqnarray}
s_{n} & = & \left(\sum_{k}\bar{u}_{k}^{n}\bar{U}_{i,j}^{k}\right)^{2}\label{eq: sn U symmetric simple}
\end{eqnarray}
This expression also shows that for symmetric $\bar{U}$, the terms
$s_{n}$ form a non-increasing sequence, and it allows to better understand
the above described behaviour of the $\mathcal{U}_{n}$ terms.\\
If $\mathcal{A}$ is close to being mutual unbiased with $\mathcal{B}$
then $\left|\bra{a_{i}}U\ket{a_{j}}\right|\approx1/\sqrt{M}\ \forall i,j$,
such that $\bar{U}_{i,j}\approx1/M$. In this case $\bar{U}\approx\ket{\bar{u}_{1}}\bra{\bar{u}_{1}}$
i.e., the single relevant eigenvalue is the largest one i.e., $\bar{u}_{1}=1$,
while for $k>1,\ \bar{u}_{k}\approx0$. Therefore $s_{n}\approx1/M\ \forall n$
and $\mathcal{U}_{1}>\mathcal{U}_{n}\ \forall n$. We thus expect
that for bases approximately mutual unbiased, the dominant term is
the Massen-Uffink one $\mathcal{U}_{1}$ for pure states or its mixed
states counter part $\mathcal{L}_{1}$.\\ If now $\mathcal{A}$ and
$\mathcal{B}$ are sufficiently far from being mutually unbiased,
then $\bar{U}\neq\ket{\bar{u}_{1}}\bra{\bar{u}_{1}}$, the eigenvalues
$\bar{u}_{k},\ k>1$ are non-vanishing and they thus become relevant
for the determination of the $s_{n}$. Since as shown in (\ref{eq: sn U symmetric simple})
$s_{n}$ is a non-increasing sequence for $n>1$, we can expect that
$s_{1}$ may become larger than $s_{n}$ for some $n\ge2$ and thus
$\mathcal{U}_{n}>\mathcal{U}_{1}$. Aside from simple cases, such
as the single qubit one developed in Section (\ref{subsec: Example I. Single qubit}),
it is in general difficult to foresee if, when and which among the
$\mathcal{U}_{n}$ for $n\ge2$ are able to provide lower bounds $\mathcal{U}_{n}>\mathcal{U}_{1}$
i.e., tighter than the Maassen-Uffink one $\mathcal{U}_{1}$. As show
in the following examples, and as confirmed by extensive simulations,
there are indeed relevant cases in which indeed $\mathcal{L}$ provides
a bound tighter that Maassen-Uffink's and other bounds available for
pure states. And thus by extension, if the bases $\mathcal{A},\mathcal{B}$
are not mutually unbiased and since $\mathcal{S}_{n}>\mathcal{S}_{1}$,
one expects that the lower bounds $\mathcal{L}_{n}$ to be tighter
than $\mathcal{L}_{1}$. 

\section{Examples\label{sec: Examples}}

In order to assess the applicability and performance of the above
outlined strategy we explicitly apply it to different examples. We
will study both the pure and the mixed state situation. For the small
dimension cases, in order to compare our results, we choose to evaluate
the lower bounds obtainable for both pure and mixed states by means
of direct sum majorization strategy \cite{Zyc_2018_Majorization},
which is one of the most sophisticated one and it has been shown to
provide the best (tighter) lower bounds in several examples with small
$M\le4$. The full method is based on the determination of an $M$-dimensional
vector 
\begin{eqnarray*}
W & = & \left(w_{1},w_{2}-w_{1},w_{3}-w_{2},...,w_{M}-w_{M-1}\right)
\end{eqnarray*}
where each of the coefficients $w_{k}$ is the largest singular value
of the sub-matrices obtainable from $U$ with $n_{r}+n_{c}=k+1$,
where $n_{r}/n_{c}$ is number of rows/columns of the sub-matrices.
Once $W$ is determined the state independent lower bound is given
by $\mathcal{L}_{Maj}=-\sum_{k}w_{k}\log w_{k}$. In case of mixed
states with spectrum $\vec{\lambda}=\left(\lambda_{1},..,\lambda_{M}\right)$
(in decreasing order), one needs to evaluate the $2M$ dimensional
vector $W^{(\lambda)}=\Lambda W$ where $\Lambda$ is an appropriate
matrix that depends on the $\lambda_{i}$'s (see \cite{Zyc_2018_Majorization}
for details). The corresponding lower bound is then given by $\mathcal{L}_{Maj}=-\sum_{k}w_{k}^{\lambda}\log w_{k}^{\lambda}$.
For small dimensions $M$ the method can be easily applied in full.
When the dimension grows exceedingly, the elements $w_{k}$ are the
result an optimization problem that becomes gradually more difficult
as $k$ increases, due to the large number $\sum_{j}^{k}\binom{M}{j}\binom{M}{k+1-j}$
of sub-matrices that one has to consider. In such case, one can determine
only the first few $k^{*}$ coefficients $w_{k}$ and use the $M$-dimensional
vectors $W_{k^{*}}=\left(w_{1},w_{2}-w_{1},..,1-w_{k^{*}-1},0,0,0..0\right)$
and $W_{k^{*}}^{\lambda}=\Lambda W_{k^{*}}$ instead. For pure states
in the high-dimensional cases we also use other formulations of the
lower bounds (see below).

As for the mixed state case, we have already mentioned that the form
in which we test our approach is given by (\ref{eq: EURs vs sum of coherences}).
This formulation is equivalent to a specific case of uncertainty relations
with memory. In this context the idea is that two parties share a
pure state $\ket{\psi}=\sum\sqrt{\lambda_{i}}\ket{i}_{1}\ket{i'}_{2}\in\mathcal{H}_{1}\otimes\mathcal{H}_{2}$,
here written in its Schmidt decomposition. Then two different measurements
$\mathcal{A},\mathcal{B}$ are applied onto $\rho_{1}=Tr_{2}[\rho_{12}]$,
while the other subsystem is used as a memory. The uncertainty of
the protocol can then be expressed in terms of the conditional entropies
\begin{eqnarray}
H\left(\mathcal{A}|2\right)+H\left(\mathcal{B}|2\right) & = & H\left(\mathcal{A}\right)+H\left(\mathcal{B}\right)-2S\left(\rho_{2}\right)\label{eq: Conditional EURs}
\end{eqnarray}
where $H\left(\mathcal{A}|2\right)=S(\rho_{A2})-S(\rho_{2})$ is the
conditional quantum entropy expressed in terms of the von Neumann
entropy of the state $\rho_{A2}=\sum_{i}\left(\ket{a_{i}}\bra{a_{i}}\otimes\mathbb{I}_{2}\right)\rho_{12}\left(\ket{a_{i}}\bra{a_{i}}\otimes\mathbb{I}_{2}\right)$
and the von Neumann entropy of the state $\rho_{2}=Tr_{1}\left[\rho\right]$.
Given the above particular setting one has that $S\left(\rho_{1}\right)=S\left(\rho_{2}\right)=S(\rho)$,
and the above expression reduces to (\ref{eq: EURs vs sum of coherences}).
Since both reduced density matrices have the same spectrum $\vec{\lambda}=\left(\lambda_{1},..,\lambda_{M}\right)$,
given by squares of the Schmidt coefficients, by changing the $\lambda_{i}$'s,
i.e., the entanglement of the bipartite state $\ket{\psi}$, one can
fix the entropy of the state $\rho$ to a desired value. Given a lower
bound for the sum $H\left(\mathcal{A}\right)+H\left(\mathcal{B}\right)$
one can then check its performance with different levels of entropy.
In the following, we use for the elements of the spectrum the expressions
$\lambda_{k}=\exp\left(\beta k\right)/\sum_{k}\exp\left(\beta k\right)$.
By changing the value of $\beta$ we can thus vary $S(\rho)\in\left[0,\log M\right]$.

\subsection{Example I. Single qubit.\label{subsec: Example I. Single qubit}}

We start by analyzing a two dimensional quantum system. While as we
now see, in this case the lower bound $\mathcal{L}$ is always less
tight that $\mathcal{L}_{Maj}$, this example, thanks to its simplicity,
allows us to review in detail the method and some of the arguments
exposed above. For a single qubit \cite{Ghirardi_Single_Qubit}, the
problem of finding a lower bound $\mathcal{L}_{B}$ can always be
reduced to the case where $\mathcal{A}=\left\{ \ket{0},\ket{1}\right\} $
is given by the eigenstates of $\sigma_{z}$ and $\mathcal{B}\left(\theta\right)$
is obtained from $\mathcal{A}$ by means of the unitary operator 
\begin{eqnarray*}
U & = & \left[\begin{array}{cc}
\cos\theta & -\sin\theta\\
\sin\theta & \cos\theta
\end{array}\right]
\end{eqnarray*}
For $\theta=0$ the two bases coincide, while for $\theta=\pi/4$
they are mutually unbiased. The unistochastic matrix $\bar{U}$ is
symmetric, and $\left(\bar{u_{1}},\bar{u}_{2}^{n}\right)=\left(1,\cos^{n}2\theta\right)$
are the positive eigenvalues of $\bar{U}^{n}$. The largest matrix
element for $\bar{U}^{n}$ is given by $\bra{0}\bar{U}^{n}\ket{0}$
or by $\bra{1}\bar{U}^{n}\ket{1}$; one has $\left|\bra{0}\left.\bar{u}_{1}\right\rangle \right|^{2}=\left|\bra{0}\left.\bar{u}_{2}\right\rangle \right|^{2}=1/2$.
Thus we can analytically compute the $s_{n}$ with (\ref{eq: sn U symmetric simple})
in terms of the eigenvalues of $\bar{U}^{n}$ and the projections
$\left|\bra{0}\left.\bar{u}_{i}\right\rangle \right|^{2}$ as 
\begin{eqnarray*}
s_{n} & = & \left[\left|\bra{0}\left.\bar{u}_{1}\right\rangle \right|^{2}\bar{u}_{1}+\left|\bra{0}\left.\bar{u}_{2}\right\rangle \right|^{2}\bar{u}_{2}^{n}\right]^{2}\\
 & = & \left(\frac{1+\cos^{n}2\theta}{2}\right)^{2}
\end{eqnarray*}
The state independent part of the lower bounds thus reads 
\begin{eqnarray*}
\mathcal{U}_{n} & = & \frac{-\log\left(\frac{1+\cos^{n}2\theta}{2}\right)^{2}}{n+1}
\end{eqnarray*}
Here it's easy to see that for $\theta>0$, as $n\rightarrow\infty$,
$\mathcal{U}_{n}\rightarrow0$. When $\theta\lessapprox\pi/4$, $\mathcal{A}$
and $\mathcal{B}\left(\theta\right)$ are quasi mutually unbiased,
the second eigenvalue \uline{$\bar{u}_{2}\approx0$ }and thus $\mathcal{U}_{1}>\mathcal{U}_{n},\ \forall n>1$.
When $\theta$ is sufficiently smaller that $\pi/4$, $\bar{u}_{2}$
and thus its contribution to $s_{n}$ becomes non-vanishing. Indeed,
by using (\ref{eq: Ration Um Un}), one finds that if $\theta$ and
$n$ are such that 
\begin{eqnarray*}
\left(\frac{1+\cos2\theta}{2}\right) & < & \left(\frac{1+\cos^{n}2\theta}{2}\right)^{\frac{2}{n+1}}
\end{eqnarray*}
then $\mathcal{U}_{n}>\mathcal{U}_{1}$ and $\mathcal{U}_{n}$ provides
a better lower bound that the Maassen-Uffink one. For example, for
$n=2$ this happens for $\theta\approx0.592$. 

\begin{figure}[h]
\subfigure{\includegraphics[scale=0.64]{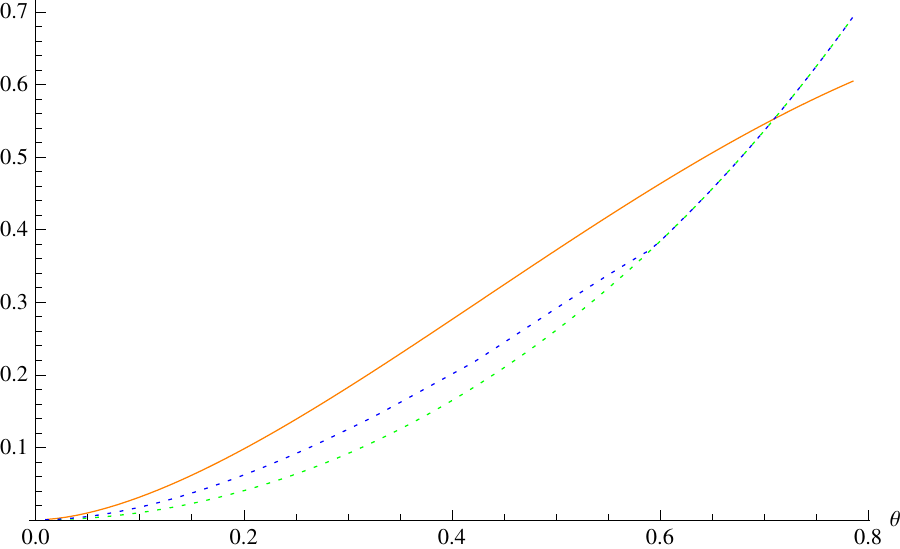} } \subfigure{
\includegraphics[scale=0.64]{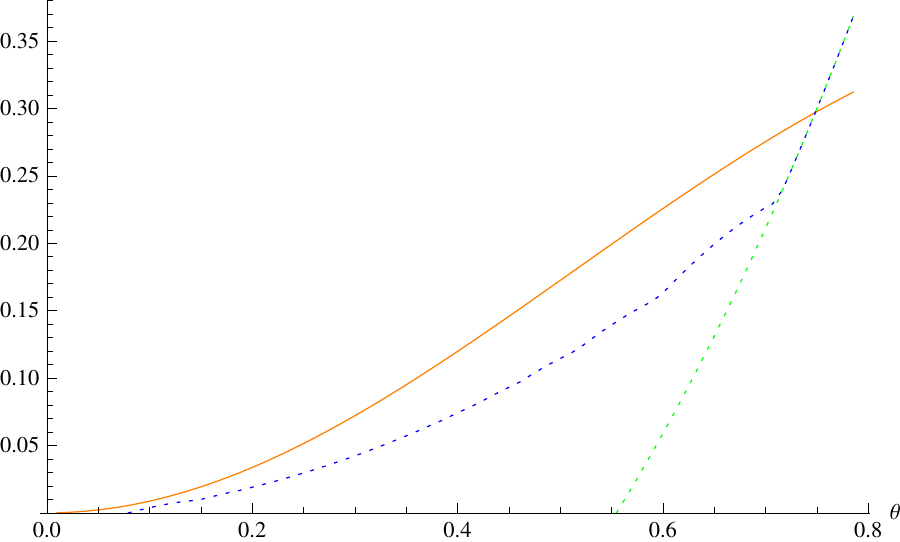}}\caption{\textbf{Single qubit. }Plots of $\mathcal{L}$ (blu dotted), $\mathcal{L}_{Maj}$
(orange),$\mathcal{L}_{1}$ (green)\textbf{ Left panel: }Pure states
case \textbf{Right panel:} Lower bound for sum of coherences (\ref{eq: EURs vs sum of coherences})
for fixed Von Neumann entropy $S\left(\rho\right)=0.32$ \label{Fig: Single Qubit}}
\end{figure}
In Figure (\ref{Fig: Single Qubit}) we plot the best lower bound
$\mathcal{L}=\max_{n}\left\{ \mathcal{L}_{n}\right\} $, together
with $\mathcal{L}_{1}$ and $\mathcal{L}_{Maj}$. In the left panel,
we show the results for the pure state case i.e., $S(\rho)=0$. For
$\theta\lessapprox\pi/4$ the two bases are mutually unbiased such
that, as expected, the best bound is provided by $\mathcal{L}_{1}=\mathcal{U}_{1}$
i.e., the Maassen-Uffink result. When $\theta$ is sufficiently smaller
that $\pi/4$ the bound provided by the direct sum majorization approach
$\mathcal{L}_{Maj}$ is always better that $\mathcal{L}$ and $\mathcal{L}_{1}$;
while, as explained above, for $\theta\le0.592$ the lower bound $\mathcal{L}$
is tighter with respect to $\mathcal{L}_{1}$.In right panel, we show
the results obtained for the expression (\ref{eq: EURs vs sum of coherences})
when $S(\rho)=0.32$. Some of the main features already discussed
above are visually reproduced by the plots. When $\theta\gtrapprox0$
the two bases $\mathcal{A}$ and $\mathcal{B}\left(\theta\right)$
are very close to each other and the dominant part of $\mathcal{L}$
is given by $\mathcal{L}_{n\ge32}\approx\mathcal{S}_{n\ge32}\approx2S\left(\rho\right)$.
Then for a large part of the interval $\theta>0$, $\mathcal{L}$
is obtained for $n>1$ and it provides a lower bound that is tighter
than $\mathcal{L}_{1}$. However, when $\theta\lessapprox\pi/2$,
$\mathcal{A}$ and $\mathcal{B}\left(\theta\right)$ become mutually
unbiased, $\bar{U}$ closely approximates the $2\times2$ van der
Waerden matrix $U^{*}$, such that $\bar{U}_{i,j}\approx1/2$. Since
$\left(U^{*}\right)^{n}=U^{*}$ then $\forall n>1\ s_{n}\approx1/4$
and consequently $\mathcal{L}_{n}\approx2\log2/\left(n+1\right)$
; thus $\forall n>1\ \mathcal{L}_{n}<\mathcal{L}_{1}\approx\log2$
and the best lower bound is given by $\mathcal{L}_{1}$.

\subsection{Example II. Three-dimensional system}

\begin{figure}[h]
\subfigure{\includegraphics[scale=0.48]{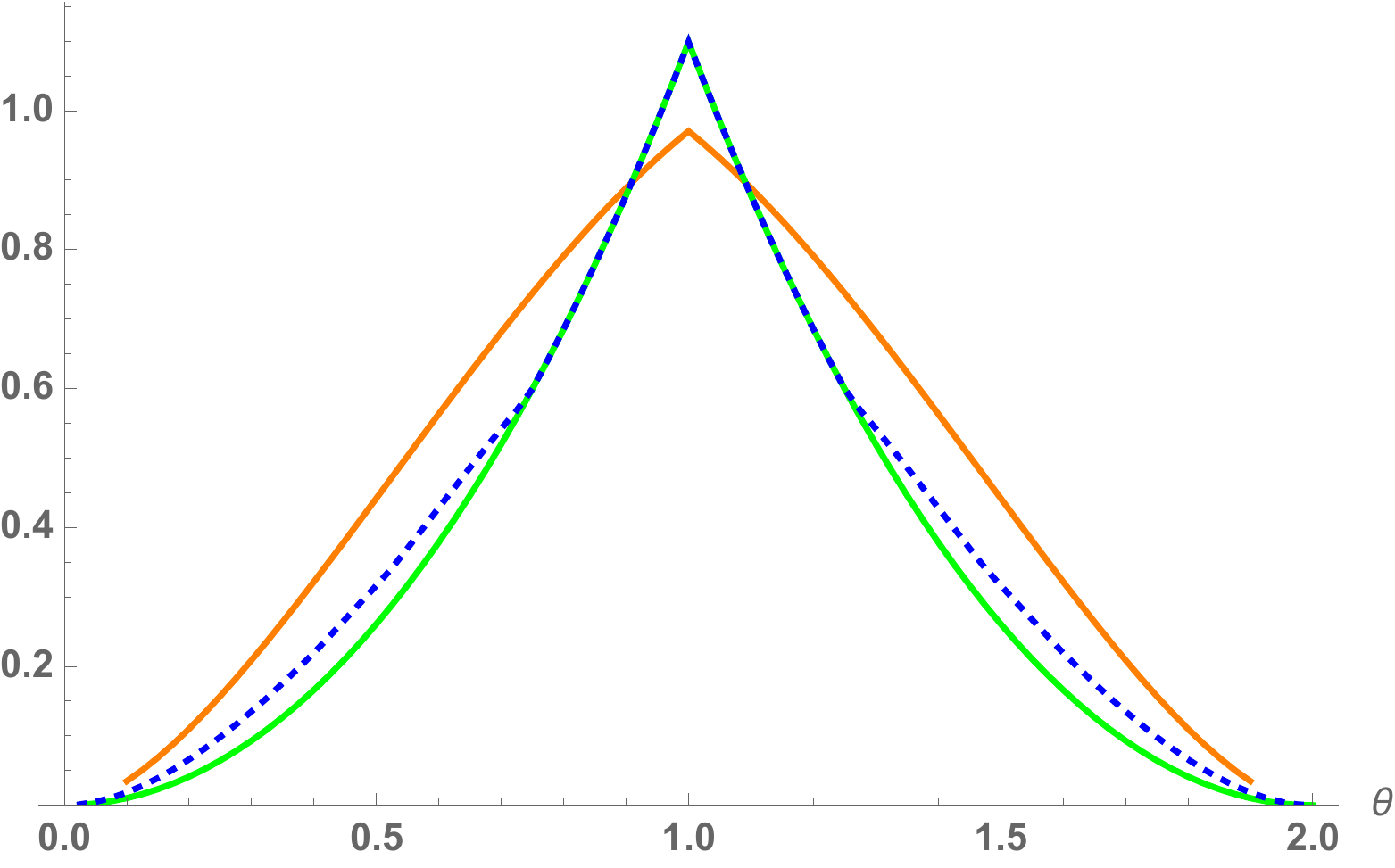}
} \subfigure{ \includegraphics[scale=0.48]{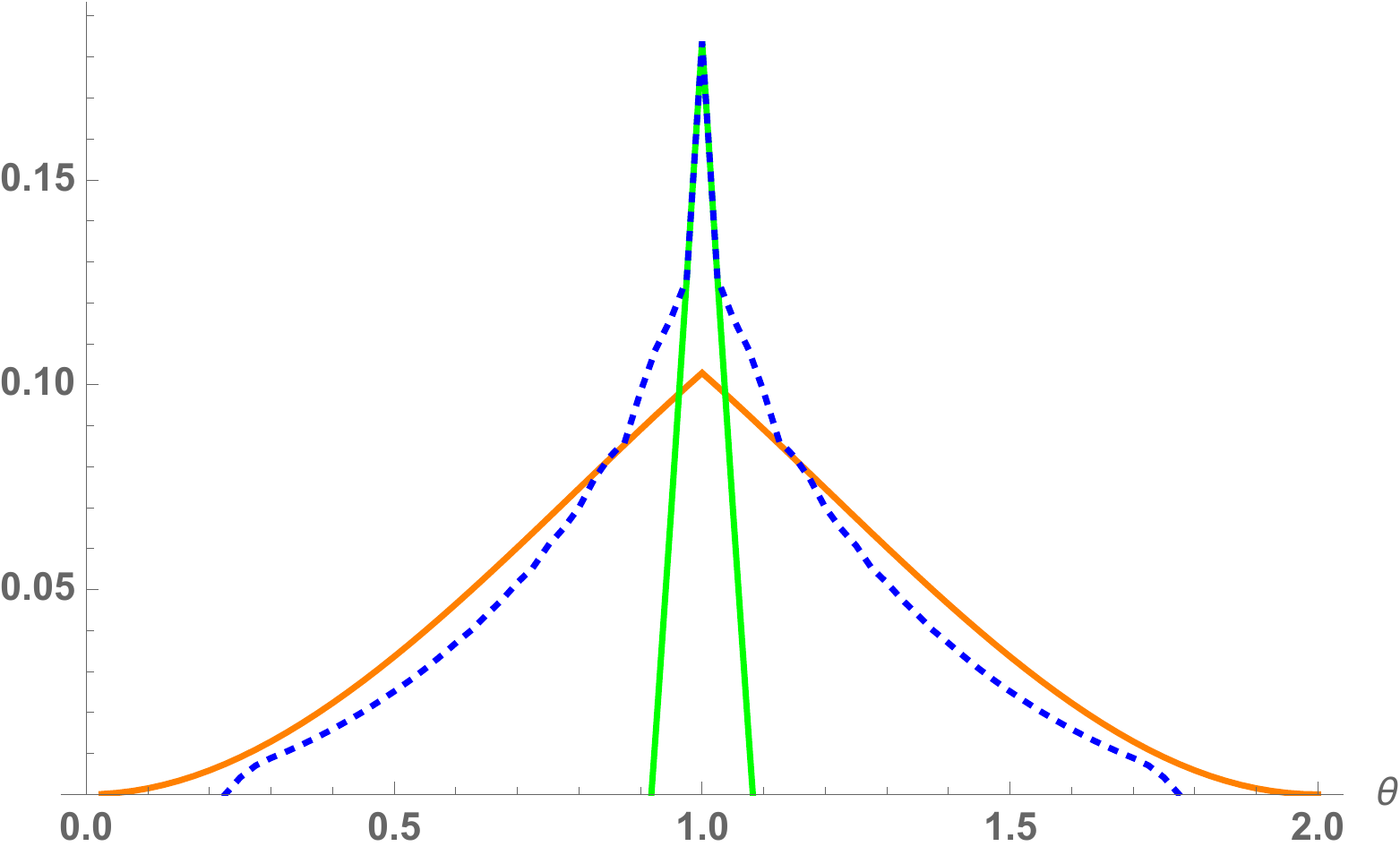}}\caption{\textbf{3-dimensional QFT. }Plots of $\mathcal{L}_{Maj}$ (orange),$\mathcal{L}_{1}$
(green),$\mathcal{L}$ (blu dotted)\textbf{ Left panel: }Pure states
case \textbf{Right panel:} Lower bound for sum of coherences (\ref{eq: EURs vs sum of coherences})
for fixed Von Neumann entropy $S\left(\rho\right)=0.914$ states case.
In a region symmetric around $\beta=1$, $\mathcal{L}$ provides a
lower bound tighter than the other ones considered.\label{Fig: QFT dim 3}}
\end{figure}

In order to test the method for three dimensional system we choose
to focus on the operator $F_{3}^{\beta}$ (also used \cite{Zyc_2018_Majorization}),
where $F_{3}$ is the three dimensional quantum Fourier transform
i.e., $\left(F_{3}\right)_{k,h}=\exp\left(2i\pi kh/3\right)/\sqrt{3}$,
and $\beta\in\left\{ 0,2\right\} $. When $\beta=1$ the bases connected
by $F_{3}$ are mutually unbiased, thus in the range $\beta\in\left\{ 0,2\right\} $
all possible values $H\left(\mathcal{A},\mathcal{B}\right)\in\left\{ 0,\ln3\right\} $
are achieved. In this case the direct sum majorization algorithm can
be fully used, and the vector $W=\left(w_{1},w_{2}-w_{1},1-w_{2}\right)$
and its counterpart for mixed states $\Lambda W$ can be easily determined.
In Fig (\ref{Fig: QFT dim 3}) left panel we plot the lower bounds
for pure states. In this case the best bound are always given by $\mathcal{L}_{Maj}$
for $\beta$ sufficiently different from $1$, and by $\mathcal{L}_{1}$
for $\beta\approx1$. The latter result, as described above, is not
surprising, since for $\beta=1$, $\mathcal{L}_{1}$ provides a tight
lower bound. For the same reason, $\mathcal{L}$ obviously coincides
with $\mathcal{L}_{1}$ for $\beta\approx1$. While away from the
central region $\mathcal{L}$ provides a slight improvement with respect
to $\mathcal{L}_{1}$. \\ The case of mixed states is described in
Fig (\ref{Fig: QFT dim 3}) right panel, where the level of the Von
Neumann entropy is fixed to $S\left(\rho\right)=0.914$. While in
general the main features discussed for pure state are again reproduced,
we notice that in a symmetric region around $\beta=1$ the bound $\mathcal{L}$
provides an advantage with respect to the other two. This result has
been confirmed by other simulations with other three dimensional operators,
such that one can induce that for very low dimensional systems, the
strategy proposed in this paper seems able to improve the existing
bounds in particular when mixed state are considered.

\subsection{Example III. High dimensional systems}

\begin{figure}[h]
\subfigure{e\includegraphics[scale=0.48]{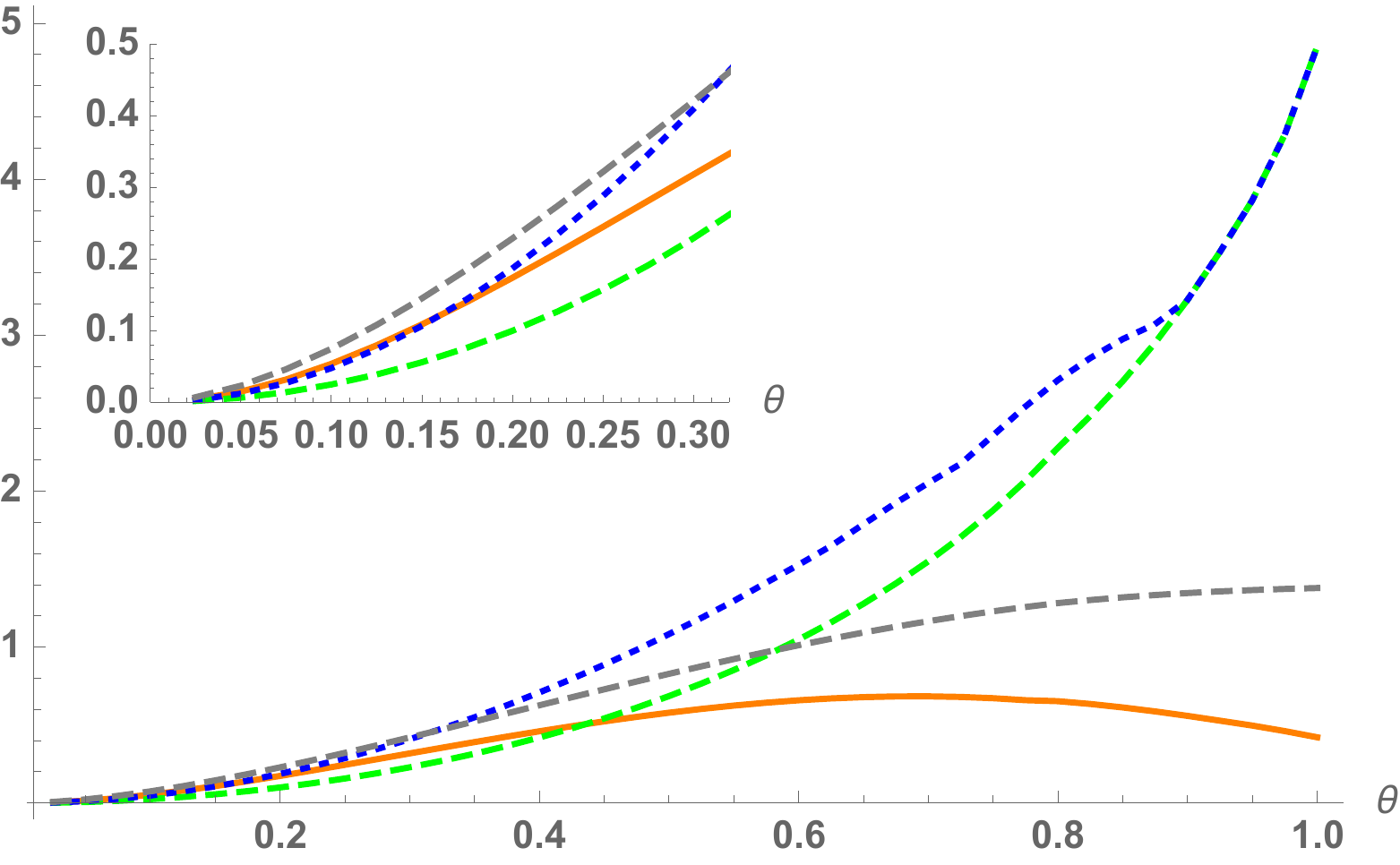}
} \subfigure{ \includegraphics[scale=0.48]{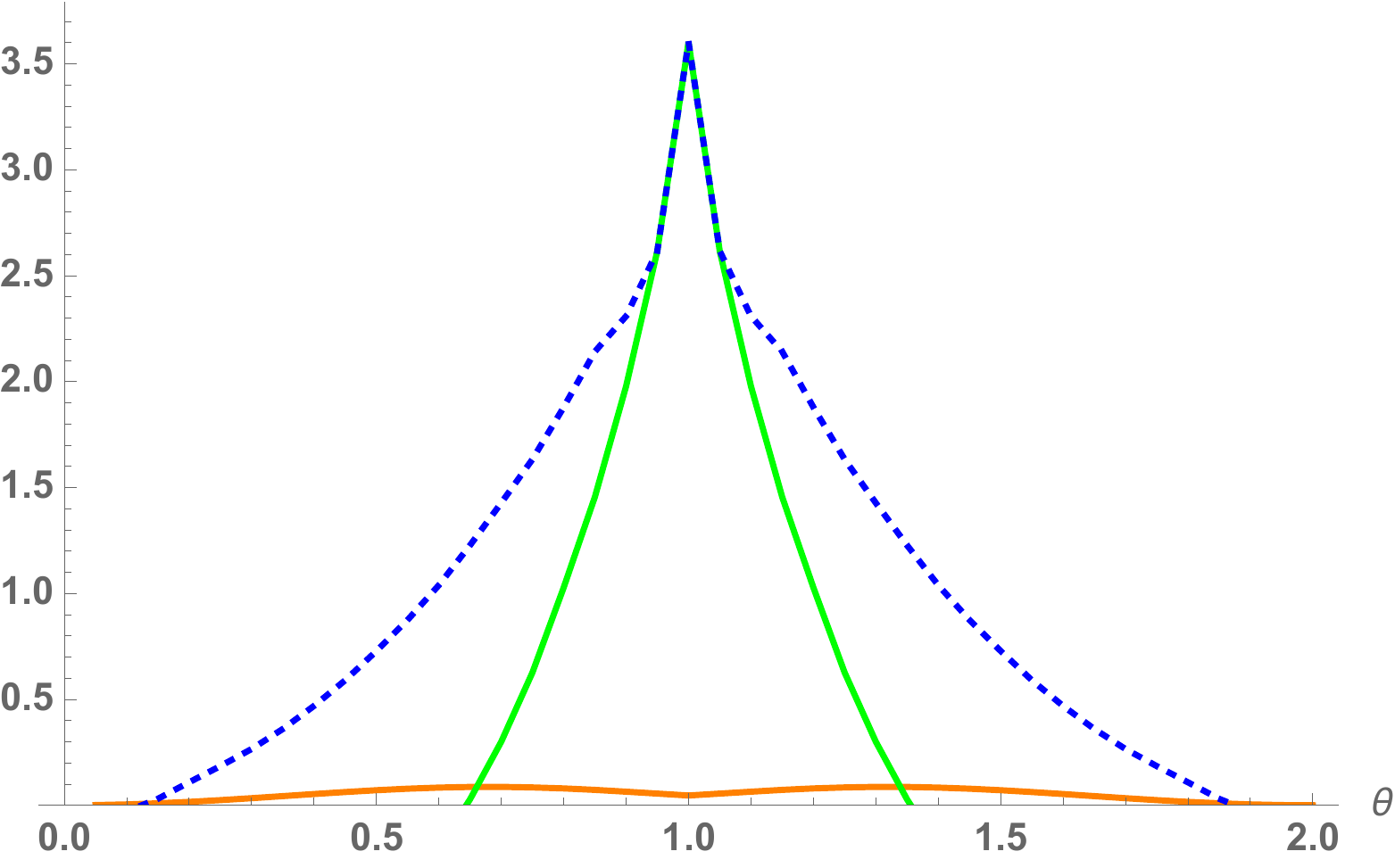}}\caption{\textbf{128-dimensional QFT. }Plots of $\mathcal{L}$ (blu dotted),
$\mathcal{L}_{Maj}$ (orange),$\mathcal{L}_{1}$ (green,dashed),$\mathcal{L}_{deV}$
(black, dashed)\textbf{ Left panel: }Pure states case \textbf{Right
panel:} Lower bound for sum of coherences (\ref{eq: EURs vs sum of coherences})
for fixed Von Neumann entropy $S\left(\rho\right)=1.25$ states case.
Here $\mathcal{L}$ provides a bound that is tighter for a large portion
of the interval considered. \label{Fig: QFT dim 128}}
\end{figure}

Being based on algorithms that have polynomial complexity in nature,
the method introduced in this work is easily applicable to situations
where other methods might be limited by their computational complexity
i.e., large dimensions. In this context, in order to test the method,
we applied it to two different operators and we fix the dimension
of the Hilbert space to $M=128$. On one hand we again use the quantum
Fourier transform operator $F_{128}^{\beta}$, and on the other hand
we use $U=\exp\left(-i2\theta J_{y}\right)$, where $J_{y}$ is the
$y$-spin operator for $j=127/2$. For high dimensions there is a
small number of lower bounds that go beyond the Maassen-Uffink result
for pure states and $\mathcal{L}_{1}$ for the mixed ones. For pure
states, there are the formulas described in \cite{ColesPianiLBsHighDim,Zyc_2014_Majorization}
\cite{deVincenteLBsHighDim} that depend on the largest element of
$\bar{U}$ and in \cite{ColesPianiLBsHighDim,Zyc_2014_Majorization}
that also depend on the second-largest matrix elements of $\bar{U}$.
For pure and mixed states we can again rely on $\mathcal{L}_{1}$
and $\mathcal{L}_{Maj}$. For the latter, as mentioned above, we opt
to use for $W$ and $W^{\lambda}$ their restricted versions $W_{2}=\left(w_{!},w_{2}-w_{1},1-w_{2},0,0,..,0\right)$
and $W_{2}^{\lambda}=\Lambda W_{2}$, that require the computation
of $w_{!},w_{2}$ only. We start by discussing the $F_{128}^{\beta}$
case. In Figure (\ref{Fig: QFT dim 128})-left panel we analyze the
pure state case, and we report $\mathcal{L}$, $\mathcal{L}_{Maj}$,
$\mathcal{L}_{1}$ and the bound $\mathcal{L}_{deV}$ given in \cite{deVincenteLBsHighDim}
that reads: 
\begin{eqnarray*}
\mathcal{L}_{deV} & = & -2\left[\left(\frac{1-\sqrt{s_{MU}}}{2}\right)\log\left(\frac{1-\sqrt{s_{MU}}}{2}\right)+\left(\frac{1+\sqrt{s_{MU}}}{2}\right)\log\left(\frac{1+\sqrt{s_{MU}}}{2}\right)\right]
\end{eqnarray*}
where again $s_{MU}$ equals the maximum matrix element of $\bar{U}$.
The other bounds \cite{ColesPianiLBsHighDim,Zyc_2014_Majorization}
are not reported since it turns out they only provide a marginal improvement
with respect to the Maassen-Uffink result $\mathcal{L}_{1}$. The
plots show that again in the region $\beta\approx1$, $\mathcal{L}_{1}$(green-dotted)
gives the best lower bounds. Away from the $\beta\approx1$, and the
newly introduced bound $\mathcal{L}$ (blue-dotted curve) is shown
to give an improvement with respect to all the other tested bounds.
In the vicinity of $\beta\gtrapprox0$, the three functionals $\mathcal{L}$,
$\mathcal{L}_{Maj}$ and $\mathcal{L}_{deV}$ give approximately the
same result, with some advantage given by $\mathcal{L}_{deV}$ (see
inset). 

In Figure (\ref{Fig: QFT dim 128})-right panel we instead consider
the mixed case scenario, and we compare $\mathcal{L}$, $\mathcal{L}_{Maj}$
and $\mathcal{L}_{1}$ when $S\left(\rho\right)=1.25$. The plot shows
that, aside from a central region where $\mathcal{L}_{1}$ (green-dotted)
dominates, for all other values the bound provided by $\mathcal{L}$
(blue-dotted) performs largely better than the other two. The same
result can be obtained with arbitrary fixed values of $S\left(\rho\right)$
.

\begin{figure}[h]
\subfigure{\includegraphics[scale=0.48]{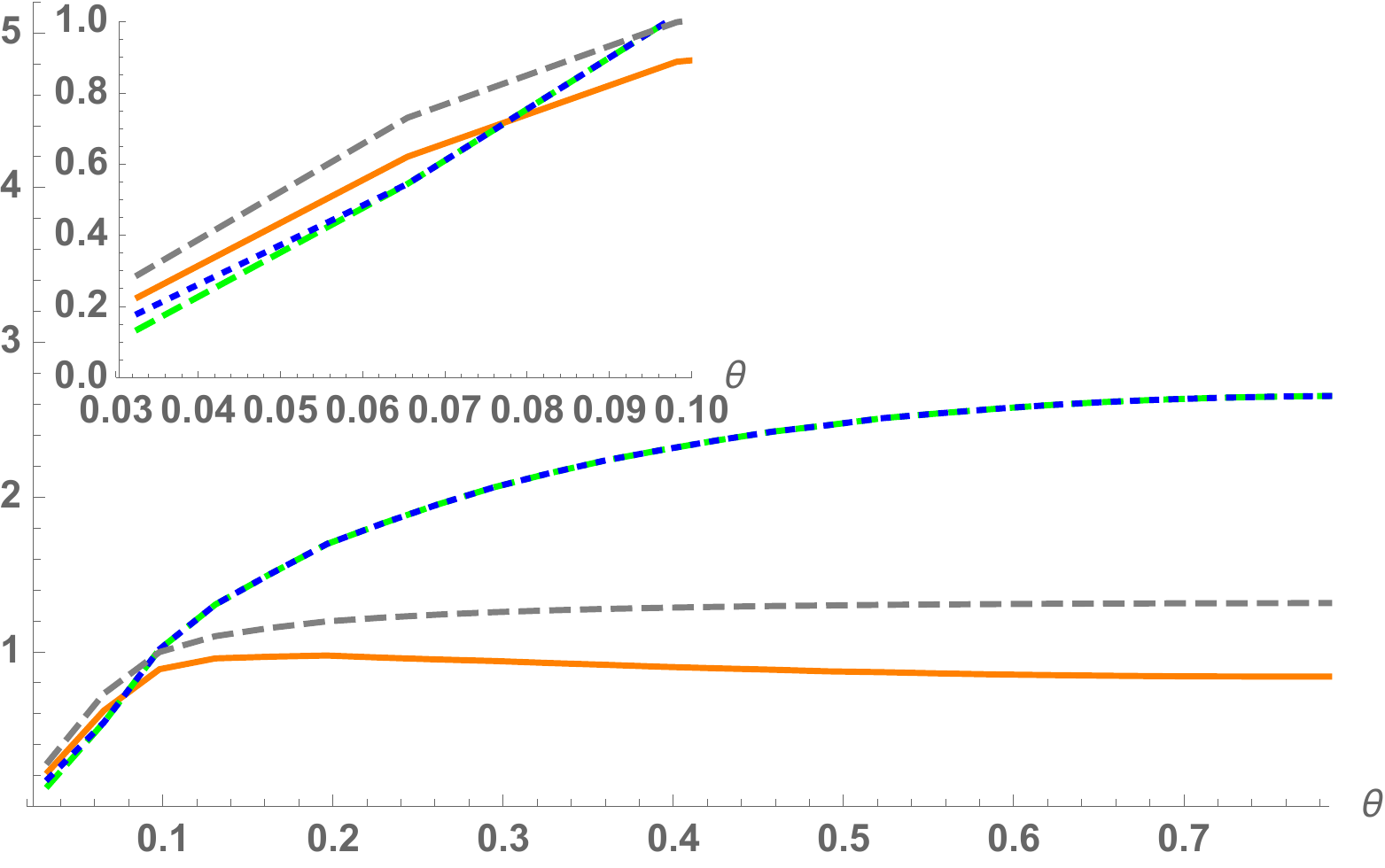}
} \subfigure{ \includegraphics[scale=0.48]{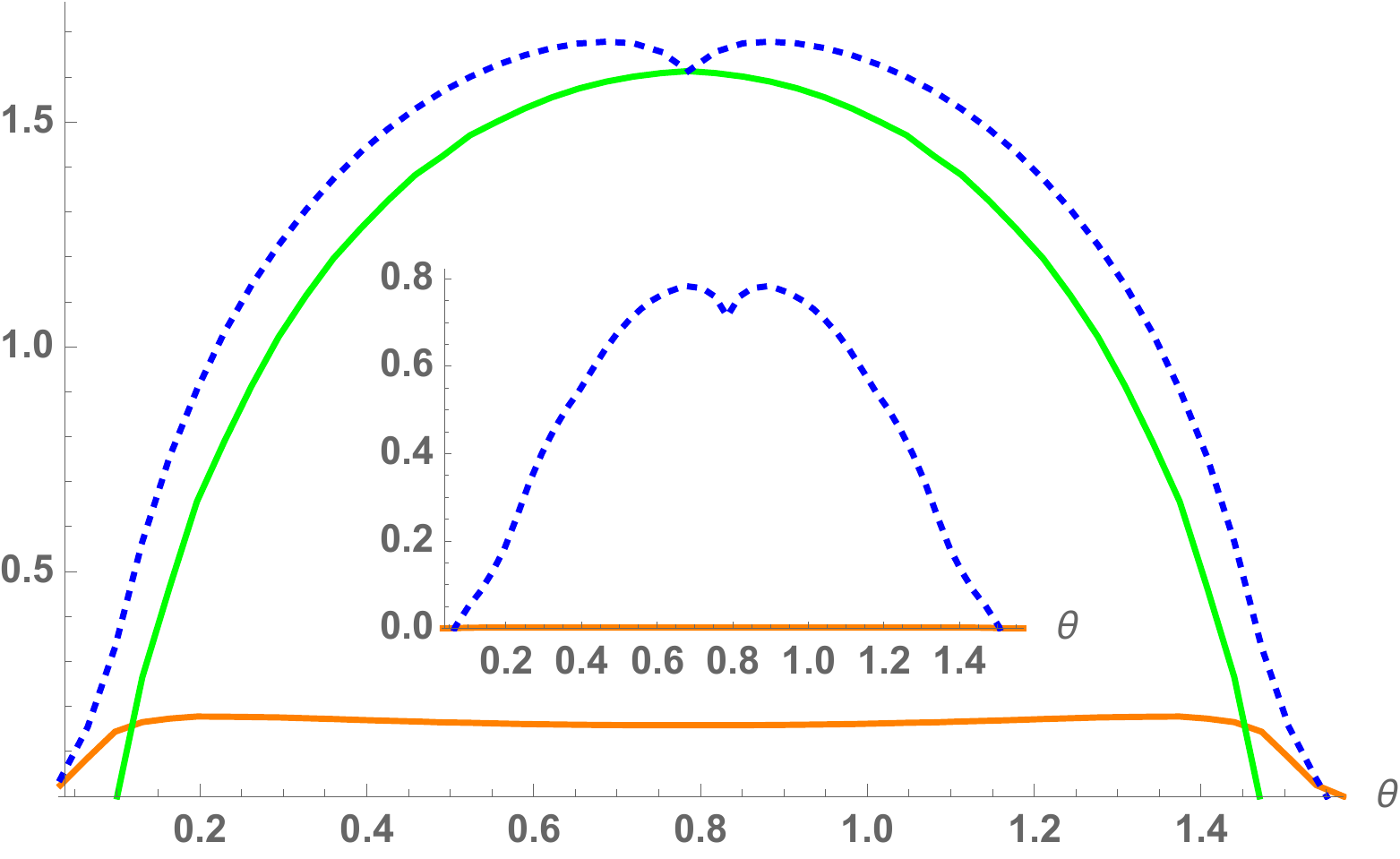}}\caption{\textbf{128-dimensional spin case. }Plots of $\mathcal{L}$ (blu dotted),
$\mathcal{L}_{Maj}$ (orange), $\mathcal{L}_{1}$ (green,dashed),$\mathcal{L}_{deV}$
(black, dashed).\textbf{ Left panel: }Pure states case \textbf{Right
panel:} Lower bound for sum of coherences (\ref{eq: EURs vs sum of coherences})
for fixed Von Neumann entropy: main plot $S\left(\rho\right)=1$;
inset $S\left(\rho\right)=1.25$ \label{Fig: Jy dim 128}}
\end{figure}

The other example is based on the unitary operator $U=\exp\left(-i2\theta J_{y}\right)$.
In (\ref{Fig: Jy dim 128})-left panel the pure state case shows that
aside from a small region around $\theta=0,\pi/2$ where $\mathcal{L}_{deV}$
and $\mathcal{L}_{Maj}$ provides some advantage (see inset for $\theta\gtrapprox0$),
for all other values of $\theta$ the best bound is given by $\mathcal{L}_{1}$.
On the other hand, for the mixed state case, starting form $S\left(\rho\right)=1$
(right panel, main plot), for all values of $\theta$, $\mathcal{L}$
is the tightest lower bound and it also provides a significant improvement
with respect to $\mathcal{L}_{1}$. When the entropy of the state
is increased i.e., $S\left(\rho\right)>1$, the improvement becomes
even more significant and for $S\left(\rho\right)\gtrapprox2.4$ (see
central inset) $\mathcal{L}$ becomes the only non-zero easily computable
lower bound available.

\section{Discussion\label{sec: Discussion}}

In this paper we have introduced a strategy to derive a sequence of
lower bounds for entropic uncertainty relations for two observables
$\mathcal{A},\mathcal{B}$. The strategy starts by recognizing that
the output probabilities of sequences of alternating measurements
onto the bases $\mathcal{A}\mbox{ and }\mathcal{B}$ can be expressed
in terms of multiple applications of bi-stochastic maps $\bar{U},\bar{U}^{T}$
that are derivable from the unitary operator $U$ connecting two bases
$\mathcal{A},\mathcal{B}$. By upper-bounding the degree of mixing
induced by the $n$-fold application of such bi-stochastic maps on
can derive lower bounds in terms of the eigenvalues and eigenvectors
of $\bar{U},\bar{U}^{T}$ that take the form $\mathcal{L}_{n}=\mathcal{U}_{n}+\mathcal{S}_{n}$.
While $\mathcal{U}_{n}$ is the state independent part, $\mathcal{S}_{n}$
only depends linearly on the Von Neumann entropy of the class of state
considered. The lower bounds obtainable with such strategy can thus
be applied to both pure and mixed states, and can be used for bounding
certain schemes involving conditional entropies and sum of coherences.
The method, being based on matrix multiplication, has a polynomial
complexity in the dimension $M$ of the underlying Hilbert space and
it thus can also be applied to high dimensional cases. We have shown
how the complexity can be further reduced, by taking advantage of
the symmetry of the operators involved. 

We have applied the method to several different examples. The main
results about the application of the method can be summarized as follows.
For the smallest case, the single qubit, the method does not provide
an appreciable advantage with respect to other existing strategies.
Already for three dimensional cases, we have shown that the method
introduced may provide some advantage in the case of mixed states.
In the high dimensional case, the results show that the method performs
better than the other scalable methods available both in the pure
and in the mixed states scenario. In the latter case the strategy
can provide non-trivial lower bounds even in the case where other
scalable methods do not.

The method can be easily extended for multiple measurement in an obvious
way i.e., by applying it to all possible pairs of bases and adding
the corresponding lower bounds. On the other hand, further research
is needed to extend the application of the method to other situations
such as for example generalized measurements (POVM).
\begin{acknowledgments}
The Author gratefully acknowledge funding from the University of Pavia
through the ``Blue sky'' project - grant n. BSR1718573 and the FRG
fund. The Author would like to thank Professor Lorenzo Maccone for
his comments to the manuscripts, fruitful discussions and his great
and precious support. 
\end{acknowledgments}

\appendix

\appendix
\numberwithin{equation}{section}

\section{Evaluation of $\mathcal{L}_{n}$\label{sec: Appendix Evaluation of L_n, Ubar non symmetric}}

We first examine the case where $\bar{U}$ is symmetric but has some
negative eigenvalues, it may happen that $s_{1}$ is provided by the
matrix element $\bar{U}_{i,j}$, while for $n=2$ by $\bar{U}_{s,t}$
with $(i,j)\neq\left(s,t\right)$. For the same reason seen in Section
\ref{subsec: Evaluation-of-the-lower-bounds}, this means that the
term corresponding to the $n$-th order is equal to 
\begin{eqnarray}
s_{n=2m+1} & = & \left(\sum_{k}\bar{u}_{k}^{2m+1}\bar{U}_{i,j}^{k}\right)^{2},\ n\mbox{ odd}\nonumber \\
s_{n=2m} & = & \left(\sum_{k}\bar{u}_{k}^{2m}\left(\bar{U}^{2}\right)_{s,t}^{k}\right)^{2}\ n\mbox{ even}\label{eq: sn even U symmetric}
\end{eqnarray}
In any case, if $\bar{U}$ is symmetric, in order to evaluate the
bounds $U_{n}$ one does not need to explicitly evaluate the powers
$\bar{U}^{n}$ in order to find the various $s_{n}$ for $n\ge2$,
but instead one only needs to find the eigenvalues and eigenvectors
of $\bar{U}$ and use the above analytical formulas.

If $\bar{U}^{T}\neq\bar{U}$ we first notice that both $\bar{U}\bar{U}^{T}$
and $\bar{U}^{T}\bar{U}$ are symmetric bi-stochastic matrices, and
thus one can readily apply the above arguments to express each $s_{n}$
for $n=2m$ even in terms of the eigenvectors and eigenvalues of $\bar{U}\bar{U}^{T}$
and $\bar{U}^{T}\bar{U}$, by using analytical formulas analogous
to (\ref{eq: sn even U symmetric}). For $n=2m+1$ odd there seem
to be no easily derivable formulas, however since the products containing
an odd number of factors will be of the form $\left(\bar{U}\bar{U}^{T}\right)^{m}\bar{U}$
or $\left(\bar{U}^{T}\bar{U}\right)^{m}\bar{U}^{T}$ one can use the
decomposition $OVO^{T}=diag\left(v_{1},v_{2},..,v_{N}\right)$ to
evaluate $OV^{n}O^{T}$ for the terms $V^{m}=\left(\bar{U}\bar{U}^{T}\right)^{m},\left(\bar{U}^{T}\bar{U}\right)^{m}$
and then evaluate the maximum matrix element of $\left(\bar{U}\bar{U}^{T}\right)^{m}\bar{U}$
and $\left(\bar{U}^{T}\bar{U}\right)^{m}\bar{U}^{T}$. Aside from
the diagonalization of $\bar{U}\bar{U}^{T}$ or $\bar{U}^{T}\bar{U}$
the procedure require two simple matrix multiplications to determine
$V^{m}$ and then the given odd term $n=2m+1$. Clearly the determination
of the odd part of the sequence $\mathcal{L}_{n}$ is indeed more
demanding from a computational point of view. However, by relying
on the even part of the sequence one can obtain a sequence of lower
bounds $\mathcal{L}_{2m}$ that are representative of the whole sequence
$\mathcal{L}_{n}$.

\textbackslash{}
\end{document}